\documentclass[12pt,letterpaper,twocolumn,british,english,sort,prl,reprint,superscriptaddress]{revtex4-1}

\usepackage[T1]{fontenc}
\usepackage[utf8x]{inputenc}
\usepackage{array}
\usepackage{textcomp}
\usepackage{amstext}
\usepackage{graphicx}
\usepackage{esint}
\usepackage{subscript}

\makeatletter

\providecommand{\tabularnewline}{\\}

\usepackage{hyperref}

\makeatother

\usepackage{babel}
\begin{document}

\title{Wrinkles of graphene on Ir(111): Macroscopic network ordering and
internal multi-lobed structure}

\selectlanguage{british}%

\author{{\small{}Marin Petrović}}

\email{mpetrovic@ifs.hr}

\selectlanguage{british}%

\affiliation{{\footnotesize{}Institut za fiziku, Bijenička 46, 10000 Zagreb, Croatia}}

\author{{\small{}Jerzy T. Sadowski}}

\affiliation{{\footnotesize{}Center for Functional Nanomaterials, Brookhaven National
Lab, Upton, NY 11973, USA}}

\author{{\small{}Antonio Šiber}}

\affiliation{{\footnotesize{}Institut za fiziku, Bijenička 46, 10000 Zagreb, Croatia}}

\selectlanguage{english}%

\affiliation{{\footnotesize{}Center of Excellence for Advanced Materials and Sensing
Devices}\foreignlanguage{british}{{\footnotesize{}, Institut za fiziku,
Bijenička 46, 10000 Zagreb, Croatia}}}

\selectlanguage{british}%

\author{{\small{}Marko Kralj}}

\affiliation{{\footnotesize{}Institut za fiziku, Bijenička 46, 10000 Zagreb, Croatia}}

\selectlanguage{english}%

\affiliation{{\footnotesize{}Center of Excellence for Advanced Materials and Sensing
Devices}\foreignlanguage{british}{{\footnotesize{}, Institut za fiziku,
Bijenička 46, 10000 Zagreb, Croatia}}}
\begin{abstract}
The large-scale production of graphene monolayer greatly relies on
epitaxial samples which often display stress-relaxation features in
the form of wrinkles. Wrinkles of graphene on Ir(111) are found to
exhibit a fairly well ordered interconnecting network which is characterized
by low-energy electron microscopy (LEEM). The high degree of quasi-hexagonal
network arrangement for the graphene aligned to the underlying substrate
can be well described as a (non-Poissonian) Voronoi partition of a
plane. The results obtained strongly suggest that the wrinkle network
is frustrated at low temperatures, retaining the order inherited from
elevated temperatures when the wrinkles interconnect in junctions
which most often join three wrinkles. Such frustration favors the
formation of multi-lobed wrinkles which are found in scanning tunneling
microscopy (STM) measurements. The existence of multiple lobes is
explained within a model accounting for the interplay of the van der
Waals attraction between graphene and iridium and bending energy of
the wrinkle. The presented study provides new insights into wrinkling
of epitaxial graphene and can be exploited to further expedite its
application.
\end{abstract}
\maketitle

\section{Introduction}

Graphene is considered to be a promising material for various technological
applications \citep{Schwierz2010,Bonaccorso2010,Brownson2011,Bae2010}
due to its superior physical properties \citep{CastroNeto2009}. However,
the desired intrinsic properties are often downgraded in chemically-synthesized
graphene, including epitaxial graphene as the major large-scale source
of the material, due to the structural defects in the honeycomb lattice
(vacancies, pentagon-heptagon defects, grain boundaries) and out-of-plane
deformations, where the latter can also be induced by the former \citep{Yazyev2014}.
Characteristic deformations in epitaxial graphene are wrinkles which
form during the synthesis due to the lateral stress induced by the
mismatch between thermal contraction of graphene and its substrate.
In the literature, the out-of-plane deformations of graphene are sometimes
also referred to as ripples, ridges or folds, depending on their size
and structural details. In the present work, we do not make this distinction
and by wrinkles we imply delaminated parts of graphene which can be
thought of as one-dimensional (1D) objects at micron-scales, while
forming a wrinkle network (WN) at larger scales as we demonstrate
in the following. Wrinkles are found in graphene on metals \citep{Sutter2009a,Chae2009,Obraztsov2007,N'Diaye2009b,Loginova2009a,Li2009,Liu2012a},
silicon carbide \citep{Biedermann2009,Sun2009,Prakash2010} as well
as various samples based on transferred graphene \citep{Lanza2013b,Zhu2012,Xu2009,Gao2012}.
Besides in graphene, wrinkling occurs in other types of layered materials
\citep{Kim2012,Kushima2014} and more generally in various systems,
ranging from semiconductors \citep{Mei2007} to soft matter and biological
tissues \citep{Li2012}.

The prominent role of wrinkles is that they affect the transport properties
of graphene resulting, for example, in tunable transport gaps \citep{Yazyev2010},
reduction and anisotropy of electrical conductivity \citep{Xu2009,Zhu2012}
and a decrease of thermal conductivity \citep{Chen2012}. Moreover,
highly curved surfaces of wrinkles exhibit increased chemical reactivity
which has been predicted earlier \citep{Srivastava1999} and was confirmed
in experiments with oxygen etching \citep{Starodub2010a,Zhang2013PCCP}
and the growth of organic semiconductor structures \citep{Khokhar2012}.
Wrinkles are also important for the process of graphene intercalation.
They facilitate the penetration of atoms across the graphene sheet
either through defective regions characteristic for wrinkles interconnection
sites \citep{Petrovic2013a,Schumacher2014} or via temporary defects
which may form due to an increased reactivity at curved graphene areas
\citep{Vlaic2014}. They also serve as pipelines for diffusion of
already intercalated species \citep{Kimouche2014,Zhang2013PCCP}.
On a macroscopic scale, wrinkles play a significant role in the wettability
and optical transmittance of graphene \citep{Zang2013}. Therefore,
a control over their macroscopic ordering accompanied by a systematic
characterization is desirable for a successful technological implementation
of graphene. Some efforts have already been made in the direction
of wrinkle manipulation, e.g. modification of wrinkle orientation
and length by the scanning tunneling microscopy (STM) tip for graphene
on silicon carbide \citep{Sun2009} or alteration of the wrinkle area
density and orientation by selecting the appropriate substrate morphology
for transferred graphene \citep{Lanza2013b,Pan2011b}.

From the published studies, typical heights and widths of wrinkles
in epitaxial graphene are found to vary from below one and up to several
tens of nanometers, depending on the substrate and graphene growth
procedure \citep{N'Diaye2009b,Loginova2009a,Liu2012a,Sun2009,Prakash2010,Zhu2012}.
Internal substructure of wrinkles is often approximated by a single-lobed,
semicircular shape \citep{Prakash2010,N'Diaye2009b,Xu2009}. However,
other physical interactions involved in the wrinkle formation (e.g.
van der Waals) may lead to the formation of more complex structures
\citep{Zhu2012,Kim2011a}.

Among epitaxially grown systems, graphene on iridium (111) surface
{[}Gr/Ir(111){]} is widely-explored due to the high structural quality
\citep{Coraux2008} and almost intrinsic electronic properties \citep{Pletikosic2009,Kralj2011}
of graphene. Up to now, experimental characterization of wrinkles
of Gr/Ir(111) via low-energy electron microscopy (LEEM) and a simple
model accounting for the wrinkle formation have been reported in the
work of N’Diaye et al. \citep{N'Diaye2009b}. There the authors focused
on graphene islands and graphene containing several rotational domains
with respect to the substrate. Similar samples were also used in the
study of Loginova et al. \citep{Loginova2009a}. Moreover, an x-ray
standing wave studies revealed correlation between graphene coverage
and its periodic buckling amplitude \citep{Busse2011a,Runte2014}
which is a consequence of the partial stress relaxation in graphene
islands via their lateral expansion. Therefore, the structure of wrinkles
is also expected to vary with graphene coverage. A detailed study
on the relation between lattice parameters of graphene and iridium
in a broad temperature interval revealed a hysteretic behavior of
wrinkle formation and their flattening \citep{Hattab2012}. The same
study showed that wrinkles on average relax $\sim2/3$ of the compressive
stress imposed by iridium while $\sim1/3$ still remains in the graphene
layer.

In this work, we are interested in wrinkles of the high-quality macroscopic
graphene samples and hence we focus on the full monolayer of Gr/Ir(111)
and refrain from the description of graphene islands and incomplete
graphene layers. LEEM is used for the characterization of WN of Gr/Ir(111).
The \textit{in situ} imaging at elevated temperatures gives a detailed
insight into the process of wrinkling. The WN is found to be well
represented by an appropriate Voronoi partition of the plane with
centroids constructed from the wrinkle positions. Analysis of STM
measurements reveals that individual wrinkles have intriguing and
complicated profiles. These are explained within a framework of a
1D model accounting for an interplay of different energies important
for the wrinkle formation. Comparison of our model with experimental
data strongly suggests that the wrinkles we observe are multi-lobed,
i.e. that the graphene approaches and delaminates from iridium several
times within the span of a single wrinkle.

\section{Experimental}

Sample preparation and all experimental measurements were carried
out \textit{in situ} under ultra-high vacuum conditions. The Ir(111)
single-crystal was cleaned by argon ion sputtering at room temperature
at 1.5 keV ion energy, followed by oxygen glowing at 850 \textdegree C
and annealing at 1200 \textdegree C. Graphene synthesis consisted
of temperature programmed growth (TPG) followed by a chemical vapor
deposition (CVD) which are known to yield full coverage of high quality
graphene monolayer of uniform orientation \citep{Coraux2009a} as
confirmed by low-energy electron diffraction (LEED) after the synthesis.
During TPG, ethylene was adsorbed on iridium at room temperature ($1.33\times10^{-7}$
mbar for 45 s), followed by a flash to 1200 \textdegree C. During
CVD, the sample was kept at 850 \textdegree C while being exposed
to ethylene ($1.33\times10^{-7}$ mbar for 900 s). The LEEM measurements
were performed at the U5UA beamline of the NSLS synchrotron (Brookhaven
National Laboratory) with an Elmitec LEEM III system. Sample temperature
($T_{s}$), imaging electron energy ($E_{i}$) and field of view (FOV)
varied and are noted in figure captions. All LEEM images were recorded
in the bright field mode. Spatial resolution of the microscope was
$\sim10$ nm. STM experiments were performed in Zagreb using Specs
Aarhus STM operating at room temperature. STM data was post-processed
in WSxM software \citep{Horcas2007}.

\section{Results and discussion}

\subsection{Wrinkle network}

\begin{figure}[t]
\begin{centering}
\includegraphics{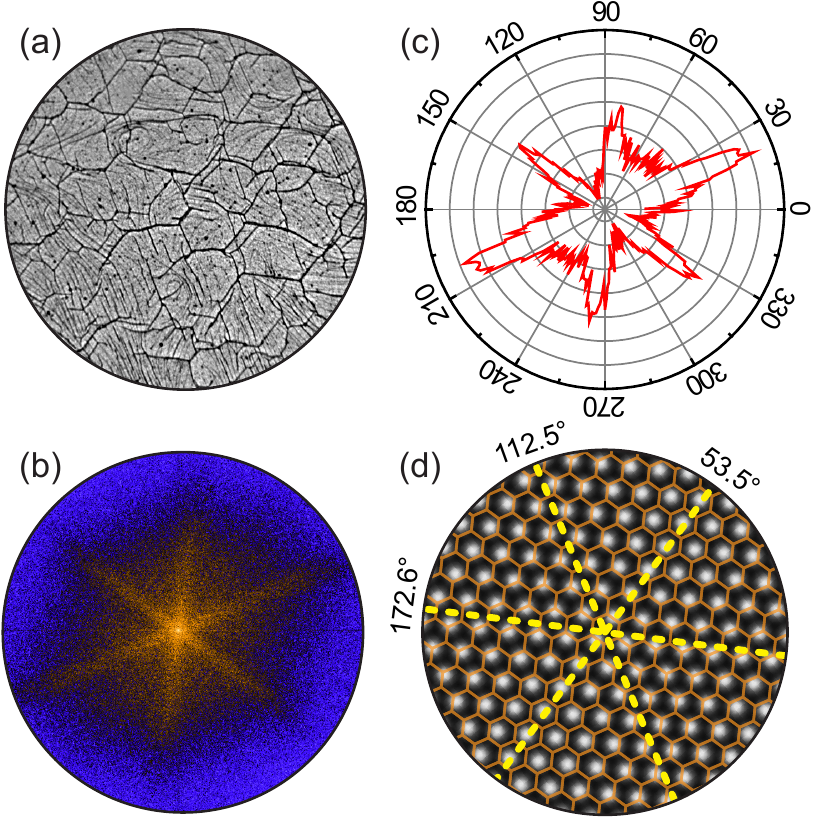} 
\par\end{centering}

\protect\caption{\textbf{\label{fig:1}}(a) LEEM image of Gr/Ir(111). Thick dark lines
extending over the entire sample surface are graphene wrinkles, thinner
wavy lines are iridium steps. FOV = 9.3 $\mu$m, $E_{i}=16$ eV, $T_{s}=96$
\textdegree C. (b) Fourier transform of image from (a). (c) Polar
plot of radial sums extracted from (b). The red curve is plotted with
a negative offset and in arbitrary units. (d) Arrangement of iridium
atoms (gray spots), graphene (orange hexagons) and dominant directions
of graphene wrinkles extension (yellow dashed lines).}
\end{figure}

Figure \ref{fig:1}a shows a LEEM image of an area of Gr/Ir(111) where
the dense-packed rows of Ir are aligned to the zig-zag directions
of graphene lattice, referred to as R0 graphene \citep{N'Diaye2008}.
Graphene wrinkles are identified as thick dark lines extending $\sim1$
$\mu$m in length and forming a network. Thinner and more closely
spaced wavy lines are steps of the iridium surface. It can be noted
that wrinkles extend mainly in three directions rotated by an angle
of $\sim60\text{\textdegree}$. As a consequence, wrinkle junctions
most often involve three wrinkles separated by 120\textdegree{} rotation.
This is directly confirmed in the Fourier transform (FT) of the background-subtracted
Figure \ref{fig:1}a, where a clear 6-fold symmetrical pattern is
revealed, as shown in Figure \ref{fig:1}b. The FT pattern consists
of stripes and lacks spots which indicates that the separation between
parallel wrinkles and their apparent thickness have a fairly broad
distribution, but the directional preference is nevertheless clear.
In order to perform a more quantitative analysis, radial sums of the
FT are presented in a polar graph in Figure \ref{fig:1}c. Again,
the 6-fold symmetry is evident. After plotting the same data in a
rectangular graph and fitting the six peaks with Gaussian curves (not
shown), one can precisely determine the main directions in which the
wrinkles extend. They are $\left(53.5\pm0.3\right)\text{\textdegree}$,
$\left(112.5\pm0.3\right)\text{\textdegree}$ and $\left(172.6\pm0.6\right)\text{\textdegree}$
(cf. yellow lines in Figure \ref{fig:1}d). The difference between
these directions fits well to 60\textdegree , which agrees with the
initial inspection of the LEEM image in Figure \ref{fig:1}a. This
kind of ordering along with the observation that wrinkle junctions
contain mostly three wrinkles suggests that the wrinkles form a network
composed dominantly of hexagons.

The exact orientation of the dense-packed atomic rows of the iridium
surface, determined by precise movement of the sample along the high-symmetry
directions during LEEM imaging, is schematically illustrated in Figure
\ref{fig:1}d. In addition, yellow dashed lines mark three determined
wrinkle-extending directions and a perfect coincidence between them
and dense-packed rows of iridium is found. The shrinkage of iridium
during cooldown after graphene synthesis can be simply described as
a reduction of separation between adjacent atomic planes in the bulk
and also atomic rows at the surface. From the observation of wrinkles
directional ordering, it can be inferred that the shrinkage induces
stress which is relaxed through the formation of wrinkles in the direction
parallel to the atomic rows of iridium i.e. parallel to the zig-zag
direction of graphene. Similar orientational locking was also proposed
earlier for graphene flakes on iridium \citep{N'Diaye2009b}.

\begin{figure}[t]
\begin{centering}
\includegraphics{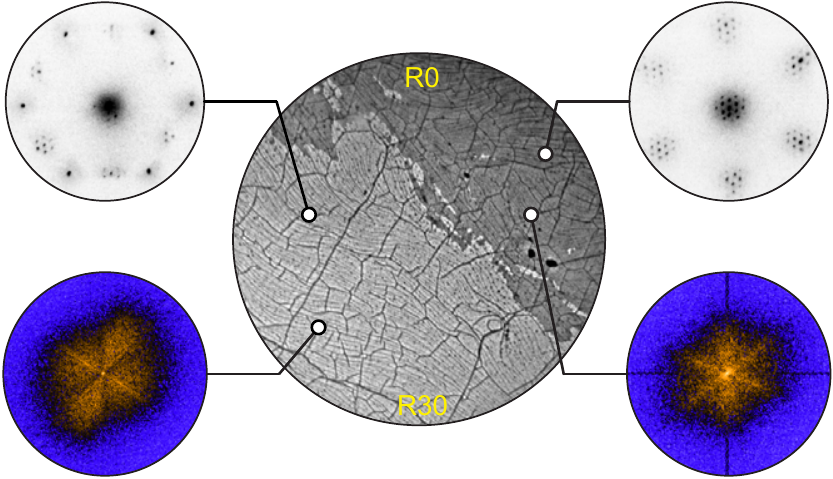} 
\par\end{centering}

\protect\caption{\textbf{\label{fig:2}}LEEM image of R0/R30 graphene boundary (center)
with the corresponding micro-LEED diffraction patterns (upper panels)
and Fourier transforms (bottom panels). FOV = 14 $\mu$m, $E_{i}=3.8$
eV, $T_{s}=98$ \textdegree C.}
\end{figure}

For a comparison to the preferred R0 graphene, orientation of the
R30 graphene wrinkles was examined where the dense-packed rows of
iridium are parallel to the armchair directions of graphene. Figure
\ref{fig:2} shows the LEEM image of a boundary area between the R0
and R30 domains of graphene, as identified via micro-LEED patterns
(upper panels), each exhibiting different LEEM reflectivity. The underlying
Ir steps and wrinkles are present on both domains. The corresponding
FT patterns from different graphene rotations are shown in bottom
panels of Figure \ref{fig:2}. Whereas the symmetry of the R0 transform
is 6-fold, the same as in Figure \ref{fig:1}b, this is not the case
for the R30 graphene where only two dominant, mutually perpendicular
directions with a large angular spread are found. In principle, two
distinct directions are enough for stress relaxation in two dimensions
and the observation of three (two) directions in the R0 (R30) graphene
calls for an explanation. We note that the graphene bending rigidity
is essentially the same along the zig-zag and armchair directions
in the range of bending radii (up to $\sim10$ nm) relevant for wrinkling
\citep{Lu2009}, ruling out the graphene lattice itself as a cause
of the observed symmetry difference. It was argued in previous studies
that the R30 graphene is more weakly bound to Ir as compared to the
R0 graphene \citep{Starodub2011}. This suggests that (i) the R0 graphene
is compelled to follow contraction of the substrate in each of the
three equivalent directions defined by the (111) surface of FCC lattice
and (ii) the R30 graphene does not have to keep track of the underlying
Ir and is free to wrinkle in two approximately perpendicular directions.
Apparently, the preference between the two wrinkling regimes is governed
by the binding strengths of R0 and R30 graphene. However, the absolute
orientation of the two wrinkling directions of R30 graphene is related
to the substrate. It is visible in Figure \ref{fig:2} that some of
the R30 wrinkles are parallel to Ir steps while others are perpendicular
to them. Similar observations were made in the case of graphene on
Pt(111), a system in which graphene is more weakly bound to the substrate
as compared to the R0 Gr/Ir(111) \citep{Sutter2009a,Zhang2013PCCP}.

\subsection{Voronoi diagram approximation}

\begin{figure}[t]
\begin{centering}
\includegraphics{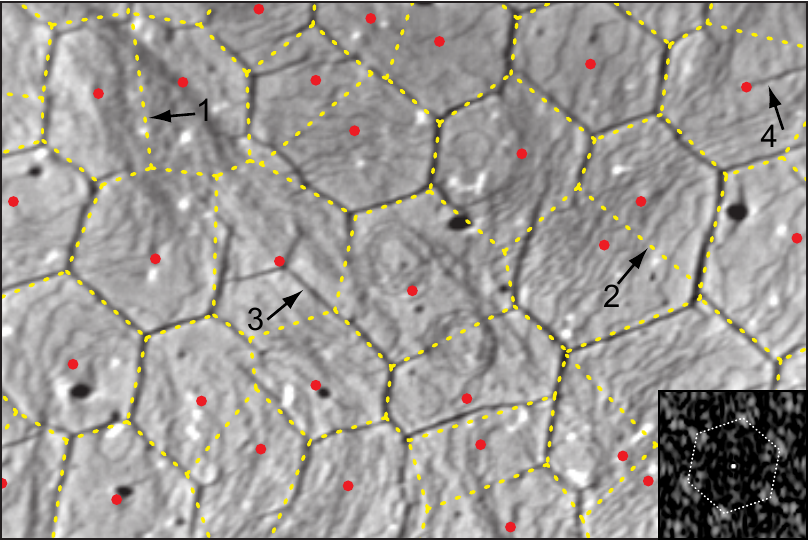} 
\par\end{centering}

\protect\caption{\textbf{\label{fig:3}}LEEM image of the graphene WN superimposed
by the VD (yellow dashed lines) generated by red centroids. Arrows
indicate several deviations of the diagram from the network. The inset
shows Fourier transform of the centroids. Image size: $\left(6.5\times4.3\right)\mu\textrm{m}^{2}$,
$E_{i}=1.9$ eV, $T_{s}=91$ \textdegree C.}
\end{figure}

On the basis of collected LEEM data, we can show that the WN on the
R0 graphene can to a good extent be approximated by a Voronoi diagram
(VD) i.e. a network obtained in a Voronoi tessellation of a plane
\citep{Okabe-book}. The construction of a VD starts from a set of
points (called centroids) which divide the plane into cells (called
Voronoi cells) associated to every centroid in a way that each cell
contains only points that are closer to a given centroid than to any
other. All Voronoi cells together constitute a VD. Figure \ref{fig:3}
shows the recorded LEEM image of Gr/Ir(111) overlapped with the VD
(yellow dashed lines) generated from a set of centroids marked as
red dots. The positions of centroids can be determined from the WN
by a simple geometrical construction applied to each cell of the network
\citep{Honda} (see Supplementary data for details). In the case of
WN, large number of wrinkles coincide with the Voronoi cell borders
with some minor exceptions. For example, the VD predicts the existence
of wrinkles which are actually not there (cf. arrows 1 and 2) or the
absence of wrinkles which are found at certain locations (cf. arrows
3 and 4). In spite of this, the VD approximation is satisfactory and
is likely to get even better in cases when various defects on the
iridium surface as well as iridium steps are suppressed to a minimum.
In addition, shorter and smaller wrinkles are hard to resolve in LEEM
and wrinkles in general can be mistaken for iridium steps and vice
versa, which may deteriorate the approximation to some small extent.

It might seem at first that the spatial distribution of centroids
in Figure \ref{fig:3} is random, but if that were the case wrinkles
would not exhibit preference in their direction and the corresponding
Fourier transforms would be isotropic, which is in contrast to what
we observe. Moreover, Fourier transform of the centroids alone, displayed
in the inset of Figure \ref{fig:3}, exhibits hexagonal symmetry and
indicates their quasi-periodic arrangement. In fact, a typical Voronoi
cell of a WN is hexagonal, as can also be concluded from the analysis
of Figure \ref{fig:1}. It follows that the WN as well as the corresponding
VD are networks of distorted hexagons similar to a honeycomb structure.
Based on these findings, WN can be simulated via construction of perfectly
arranged honeycomb VD to which a specific amount of disorder has been
introduced (see Figure S2 in the Supplementary data). VDs generated
from a random distribution of centroids are referred to as Poisson-Voronoi
diagrams and have been studied extensively \citep{Hilhorsta2008,Okabe-book}.
Expected values of various parameters (cell area, cell perimeter,
number of edges etc.) of such diagrams have been calculated and simulated.
These values are not expected to vary significantly for centroid distributions
other than random \citep{Okabe-book}, so they can be adopted for
the characterization of the WN found for wrinkles in the R0 graphene.
It is worth noting that the expected value of an angle at a typical
wrinkle junction is 120\textdegree , as we also inferred from the
data presented in Figure \ref{fig:1}. One of the easiest quantities
to obtain from LEEM is the number of wrinkle junctions per unit area,
$n_{j}$. Once this number is known, expected values of several parameters
of the network can be calculated from the theory of Poisson-Voronoi
tessellation \citep{Okabe-book}. For the LEEM image in Figure \ref{fig:3},
one finds $n_{j}=1.4$ $\mu$m\textsuperscript{-2} and other derived
parameters are given in Table \ref{tab:1}. For comparison, we also
list the values of other parameters obtained from the experiment (Figure
\ref{fig:3}) which show good agreement with the theoretical predictions.
Hence, by identifying the WN as a VD, one can easily estimate various
parameters relevant for network’s general description which in turn
facilitates evaluation of the effect the wrinkles may have on the
properties of graphene, e.g. electronic transport. Finally, the values
given in the table are not universal but are expected to differ depending
on the preparation parameters of graphene, especially the growth temperature.

\begin{table*}
\protect\caption{\label{tab:1}Parametrization of the WN. Parameters of the Poisson-Voronoi
diagram (1\protect\textsuperscript{st} column), their mutual dependence
(2\protect\textsuperscript{nd} column), values extracted from the
LEEM image shown in Figure \ref{fig:3} (3\protect\textsuperscript{rd}
column) and values calculated from the equations given in the second
column (4\protect\textsuperscript{th} column). Calculations were
performed based on the experimentally determined value $n_{j}=1.4$
$\mu$m\protect\textsuperscript{-2}.}

\centering{}%
\begin{tabular}{|>{\centering}m{4.5cm}|>{\centering}m{2.8cm}|>{\centering}m{2.8cm}|>{\centering}m{2.8cm}|}
\hline 
parameter  & relation to $n_{j}$  & experimental value  & calculated value\tabularnewline
\hline 
{\footnotesize{}density of junctions, $n_{j}$}  & $n_{j}$  & 1.4 $\mu\textrm{m}^{-2}$  & --\tabularnewline
{\footnotesize{}density of centroids}  & $\frac{n_{j}}{2}$  & --  & 0.7 $\mu\textrm{m}^{-2}$\tabularnewline
{\footnotesize{}wrinkle length per unit area}  & $2\sqrt{\frac{n_{j}}{2}}$  & 1.4 $\mu$m$^{-1}$  & 1.7 $\mu\textrm{m}^{-1}$\tabularnewline
{\footnotesize{}area of a cell}  & $\frac{2}{n_{j}}$  & --  & 1.5 $\mu\textrm{m}^{2}$\tabularnewline
{\footnotesize{}perimeter of a cell}  & $4\sqrt{\frac{2}{n_{j}}}$  & --  & 4.8 $\mu\textrm{m}$\tabularnewline
{\footnotesize{}length of a single wrinkle}  & $\frac{2}{3}\sqrt{\frac{2}{n_{j}}}$  & $\left(0.7\pm0.3\right)$ $\mu$m  & 0.8 $\mu\textrm{m}$\tabularnewline
\hline 
\end{tabular}
\end{table*}

\subsection{Dynamics of wrinkling}

\begin{figure}[t]
\begin{centering}
\includegraphics{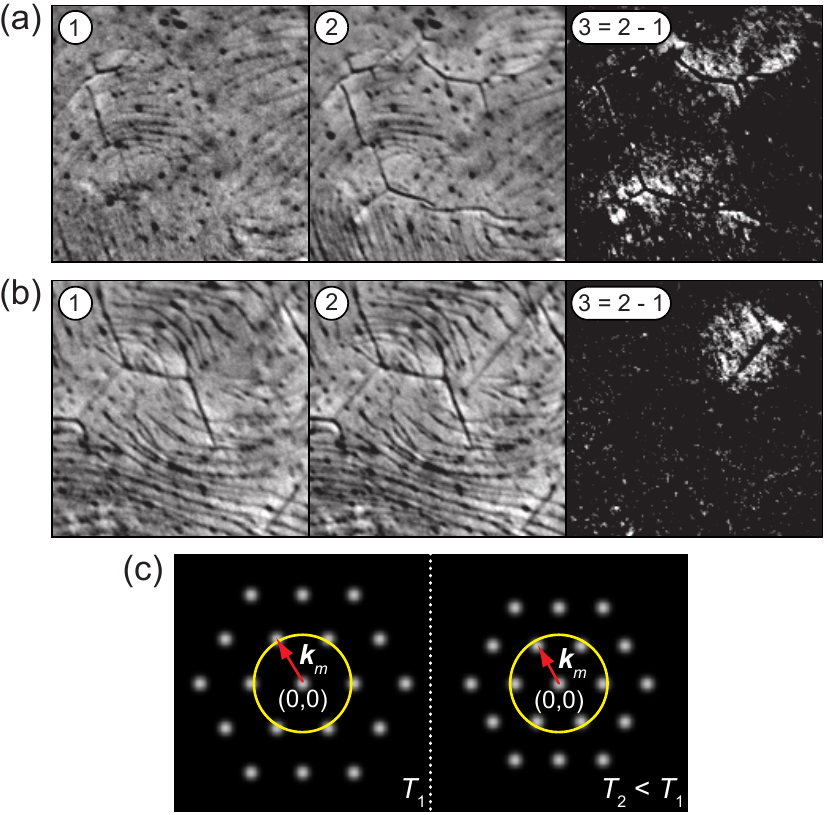} 
\par\end{centering}

\protect\caption{\label{fig:4}(a, b) LEEM micrographs, ($3\times3$) $\mu\textrm{m}^{2}$
in size, showing two different areas in of Gr/Ir(111) right before
(frame 1) and after (frame 2) wrinkle emergence. Frame 3 displays
the difference $2-1$. (a) is recorded at 350 \textdegree C, (b) at
270 \textdegree C. $E_{i}=3.8$ eV. (c) An illustration of the origin
of LEEM reflectivity changes upon wrinkle formation (see text for
details). White spots represent $\left(0,0\right)$ and moiré diffraction
spots. Yellow circle indicates size of the contrast aperture used
for LEEM imaging.\textbf{ }$\mathbf{k}_{m}$ is the moiré wave vector,
its variation at temperatures $T_{1}$ and $T_{2}$ is exaggerated
for clarity.}
\end{figure}

After the formation of 1 ML graphene and cooldown to room temperature,
the sample was re-heated to 650 \textdegree C and then slowly cooled
down while LEEM images were being simultaneously recorded. Such experiment
enabled the observation of disappearance and reappearance of individual
wrinkles. In Figure \ref{fig:4}a and b we show two areas of the sample
recorded right before (frame 1) and after (frame 2) the appearance
of several wrinkles during cooldown. By comparing frames 1 and 2,
we deduce that the reflectivity of graphene surrounding the appearing
wrinkles increases by $3\%$ once the wrinkle is formed (see also
the movie clip in the Supplementary data). This is even clearer in
frame 3 where the difference of frames 2 and 1 is shown. This effect
is discussed in the work of N'Diaye et al. \citep{N'Diaye2009b} to
be a result of a change in interference conditions of incoming and
reflected electron waves. However, in the following we propose an
additional mechanism for this local change of reflectivity.

In our experiment, LEEM images are recorded in bright field mode at
3.8 eV electron energy and at such low energies only the (0,0) spot
and a fraction of higher order moiré spots contribute to image formation
(cf. Figure \ref{fig:4}c). Furthermore, at the onset of wrinkle formation,
the surrounding graphene lattice relaxes, leading to an increase of
the graphene lattice parameter and consequently a decrease of the
moiré wave vector module $k_{m}$. This brings moiré diffraction spots
closer to the (0,0) spot by roughly $3\%$ \citep{Hattab2012} (cf.
Figure \ref{fig:4}c). Hence, a decrease of $k_{m}$ results in an
increase of the fraction of moiré spots intensity contributing to
LEEM images. In other words, areas in LEEM images which become brighter
after a wrinkle forms are areas where the graphene lattice is relaxed
as compared to its surroundings. In this way graphene lattice relaxation
can be directly observed in real space. This also indicates that the
stress imposed on graphene is visibly relaxed only in the $\sim0.5$
$\mu$m vicinity of the wrinkle and that graphene areas situated further
away remain highly stressed, at least at elevated temperatures of
several hundred degrees Celsius. The scenario of localized stress
relaxation is supported by recent theoretical simulations showing
that graphene stretching energy is reduced in the wrinkles' surroundings
\citep{Zhang2014}. Upon further sample cooldown, the reflectivity
(and hence graphene stress) in the entire LEEM field of view equalizes
but still exhibits local variations which appear to be related to
iridium surface steps. Similar relation between stress non-uniformity
and substrate morphology has been observed for graphene on silicon
carbide \citep{Robinson2009}. Finally, at room temperature when the
surface is characterized by a fully formed WN, the reflectivity variations
are diminished and indicate uniformity of stress in graphene.

\begin{figure*}
\begin{centering}
\includegraphics{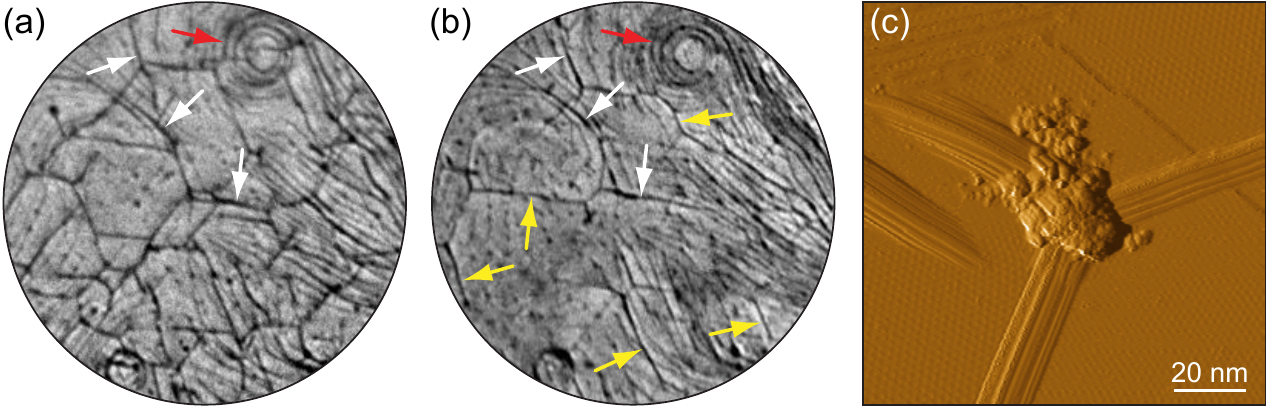} 
\par\end{centering}

\protect\caption{\label{fig:5}(a, b) LEEM images illustrating wrinkle pinning. Several
graphene wrinkles are visible (a) at 247 \textdegree C during heating
and (b) at 247 \textdegree C during cooling. Some of the wrinkles
reappeared in the same locations (white arrows) while some formed
in new locations (yellow arrows). Characteristic Ir terraces (red
arrows) facilitate identification of the same surface area of the
sample. FOV = 7.2 $\mu$m, $E_{i}=3.8$ eV. (c) STM image of the wrinkle
junction anchored at leftover dirt particle (first derivative of the
topography image). Tunneling voltage: 630 mV, tunneling current: 1.7
nA.}
\end{figure*}

Monitoring of the disappearance and reappearance of wrinkles in LEEM
during re-heating enabled examination of wrinkle nucleation sites.
Figures \ref{fig:5}(a) and (b) show two LEEM images of the same area
on the sample. Image (a) was recorded at 247 \textdegree C during
heating and image (b) at 247 \textdegree C during cooling of the sample.
In between, the sample temperature reached a maximum of 650 \textdegree C
when all wrinkles were flattened out. Re-heating cycles involve thermal
drift of the sample, but prominent iridium features (such as a series
of distinct circular terraces, cf. red arrow in Figure \ref{fig:5})
enabled identification and observation of the same surface area for
the given temperature. Comparison of the two images allows us to infer
that some of the wrinkles reappeared at the same location (cf. white
arrows in Figure \ref{fig:5}) but many other formed at new positions
which were wrinkle-free before re-heating (some of them marked by
yellow arrows in Figure \ref{fig:5}). The number of wrinkles in Figure
\ref{fig:5}b is apparently lower than in Figure \ref{fig:5}a which
is due to the hysteretic behavior of the wrinkle formation \citep{Hattab2012}.
Additional wrinkles appear as the sample is further cooled down to
room temperature, but due to mentioned thermal drift a more precise
comparison for exactly the same surface area cannot be made. We also
observe that new winkles often form at the endpoints of already existing
ones and the whole process sometimes resembles to \textquotedbl{}crack\textquotedbl{}
(wrinkle) propagation (see also the movie clip in the Supplementary
data).

We conclude from our experiments that various structural features
of graphene or iridium can anchor some of the wrinkles, serving as
nucleation sites for their formation. However, many wrinkles are exempt
form this. A typical wrinkle junction containing dirt particle, most
likely a leftover from the crystal cleaning process, was characterized
with STM and is shown in Figure \ref{fig:5}c. Such sites are suitable
for the nucleation of wrinkles due to the reduced binding of graphene
to the substrate and a presumably large number of defects contained
in the graphene lattice itself. If dirt particles are stable at elevated
temperatures, they can anchor wrinkles i.e. they are preferred sites
of wrinkle nucleation. This is in line with previous findings on graphene
flakes where the authors claim that majority of wrinkles nucleate
at defects, more precisely at heptagon-pentagon pairs \citep{N'Diaye2009b}.

\subsection{Multi-lobed wrinkles}

\begin{figure*}
\begin{centering}
\includegraphics{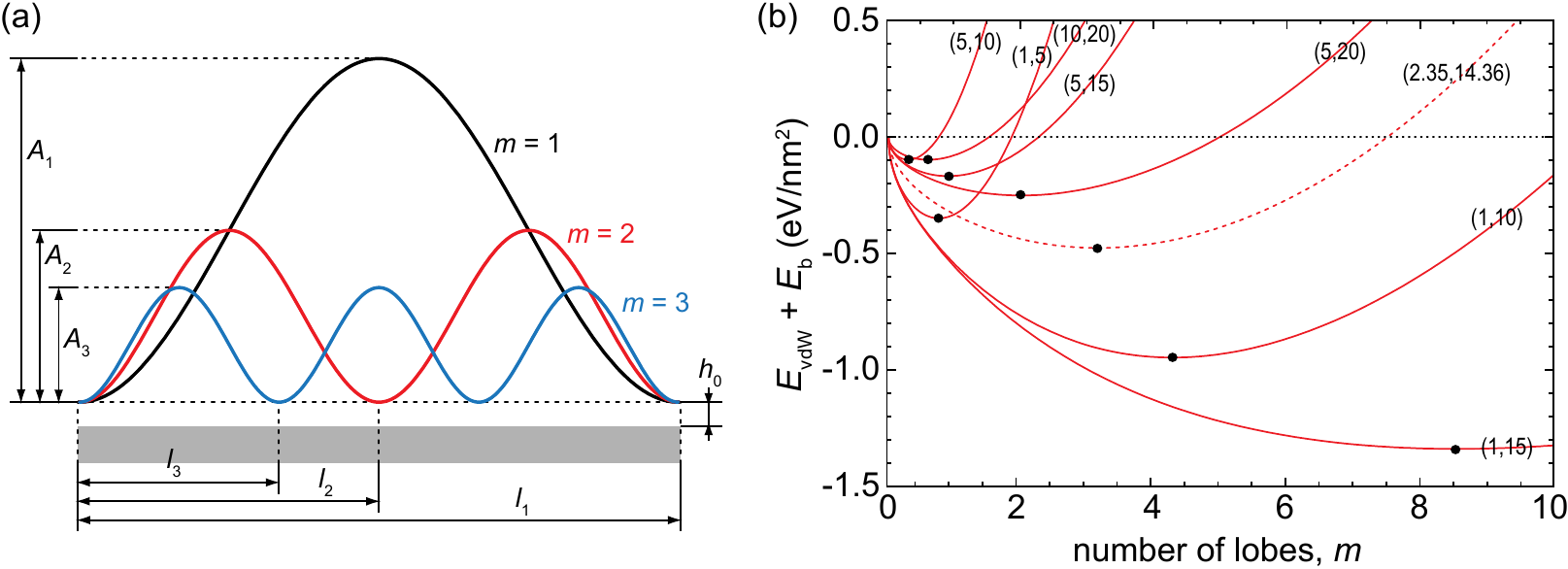} 
\par\end{centering}

\protect\caption{\label{fig:6}(a) Cross-sections of multi-lobed wrinkles for $m=1,2$
and 3 (see text for details). $h_{0}$ is the equilibrium graphene-substrate
separation. (b) The sum of binding and bending energies of a wrinkle
per unit area as a function of number of lobes within a wrinkle. Energy
minima are found for different number of lobes, as indicated by black
dots. The numbers in brackets correspond to $\left(A_{1},l_{1}\right)=\left(mA_{m},ml_{m}\right)$
in nanometers and optimal $m$ is determined for each curve from the
position of its minimum. The red dashed curve is plotted in accordance
with the parameters obtained from the STM data shown in Figure \ref{fig:7}.}
\end{figure*}

\begin{figure}[t]
\begin{centering}
\includegraphics{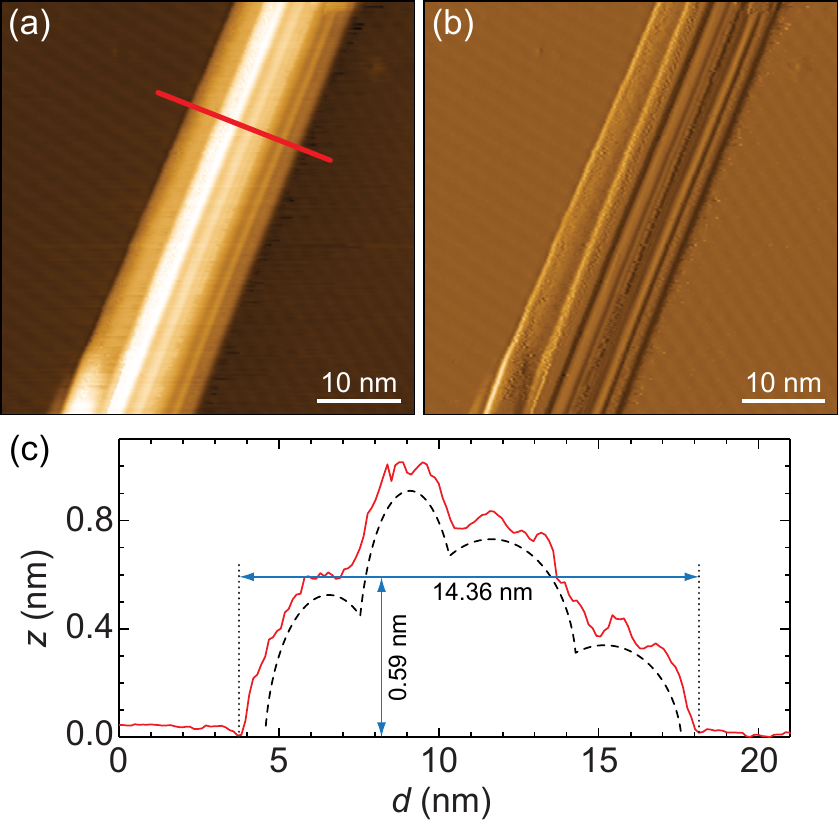} 
\par\end{centering}

\protect\caption{\label{fig:7}STM measurements of multi-lobed wrinkle. Topography
is shown in (a) and the corresponding first derivative is shown in
(b). Panel (c) shows a profile across the red line in (a) where four
lobes are identified as schematically indicated by black dotted line.
Blue arrows mark total width and average height of the wrinkle. Tunneling
voltage: 630 mV, tunneling current: 2 nA.}
\end{figure}

We now proceed to examine the features of individual wrinkles, in
particular their widths and heights and the variation of graphene-iridium
separation within a single wrinkle, i.e. the wrinkle profile. A simple,
single-lobed profile is often considered, but one may examine more
complicated profiles consisting of several consecutive lobes. Motivation
for such profiles is found in the STM measurements (shown in Figure
\ref{fig:5}c and below, in Figure \ref{fig:7}) and it is of interest
to study the energetics of their formation. The energies of different
profiles are compared using a simple \textit{ansatz} for the profile
shape which has been shown to accurately describe the compression-governed
delamination of graphene \citep{Zhang2013} and other thin films \citep{Cleymand2001,Vella2009}.
The profile is represented by the cosine function of the amplitude
$A_{m}$ extending over $m$ periods of length $l_{m}$, leading thus
to $m$ successive identical lobes. The total width of the profile
is $l_{1}$ (the width of a single-lobed wrinkle) for every $m$.
The out-of-plane displacement of $m$-lobed wrinkle in the interval
$x\in\left[-l_{1}/2,l_{1}/2\right]$ can be expressed as

\begin{equation}
w_{m}\left(x\right)=\frac{A_{m}}{2}\left[1+\cos\left(\frac{2\pi x}{l_{m}}-q\pi\right)\right]
\end{equation}

\noindent where $q$ is 0 if $m$ is odd and is 1 if $m$ is even.
An example is given in Figure \ref{fig:6}a where besides a basic,
single-lobed wrinkle ($m=1$, black line) we also show wrinkles having
two ($m=2$, red line) and three ($m=3$, blue line) lobes. In order
to be able to directly compare wrinkles containing different numbers
of lobes, the total length of wrinkled graphene in the model is kept
constant. This condition in turn yields the periods and amplitudes
of multi-lobed wrinkles: $l_{m}=l_{1}/m$ and $A_{m}=A_{1}/m$ (cf.
Figure \ref{fig:6}a, see Supplementary data for details).

\noindent 

As the total delaminated graphene length in the model is the same
in all profiles, the strain within graphene and the accompanying strain
energy do not change for different profiles. Thus, the profitability
of multiple lobes depends only on the balance between the unfavorable
bending energy cost and the favorable van der Waals binding energy
gained in the areas where graphene is brought close to iridium. In
principle, additional energy can also be gained from the van der Waals
interaction between the neighboring lobes \citep{Zhu2012}. However,
for typical profiles considered here, this contribution in negligible
and is not taken into account. We model the van der Waals energy of
the profile by summing the Lenard-Jones 6-12 interactions between
the atoms in the graphene and iridium, so that the binding energy
per unit area of the profile is given by \citep{Aitken2010}

\begin{widetext}

\begin{equation}
E_{vdW}=-\frac{\gamma}{l_{1}}\intop_{-l_{1}/2}^{l_{1}/2}\left[\frac{3}{2}\left(\frac{h_{0}}{w_{m}(x)+h_{0}}\right)^{3}-\frac{1}{2}\left(\frac{h_{0}}{w_{m}(x)+h_{0}}\right)^{9}\right]dx,\label{eq:Ebi}
\end{equation}

\end{widetext}

\noindent where $h_{0}$ is the equilibrium graphene-iridium separation
which is set to 3.4 Å \citep{Busse2011a}, and $\gamma$ is the graphene-iridium
(111) binding energy per unit area of a flat graphene; it amounts
to 50 meV per C atom (0.308 J/m\textsuperscript{2}) \citep{Busse2011a}.
The bending energy per unit area is given by \citep{Timoshenko-book}
\begin{equation}
E_{b}=\frac{D}{2l_{1}}\intop_{-l_{1}/2}^{l_{1}/2}\left(\frac{d^{2}w_{m}(x)}{dx^{2}}\right)^{2}dx,\label{eq:EBe}
\end{equation}
where $D$ is the graphene bending rigidity having a value of 0.238
nN$\cdot$nm \citep{Lu2009,Aitken2010,Zhang2013}. The comparison
of the total energy $E_{vdW}+E_{b}$ for different number of lobes
$m$ is shown in Figure \ref{fig:6}b for several combination of the
parameters $\left(A_{1},l_{1}\right)=\left(mA_{m},ml_{m}\right)$
(additional details of the calculation are given in the Supplementary
data). Black dots mark the minima of the displayed curves. The graph
shows that for certain combinations of $l_{1}$ and $A_{1}$ it is
energetically more favorable to form multi-lobed wrinkles. For example,
a single-lobed wrinkle with $A_{1}=5$ nm and $l_{1}=20$ nm {[}cf.
$\left(5,20\right)$ curve in Figure \ref{fig:6}b{]} is less preferable
than a wrinkle containing the same amount of graphene but having two
lobes so that its amplitude and wavelength are $A_{2}=A_{1}/2=2.5$
nm and $l_{2}=l_{1}/2=10$ nm, respectively. In general, multi-lobed
wrinkles become preferable when the width of a wrinkle is much larger
than its amplitude. In such cases, energy gained by additional binding
of graphene to iridium upon formation of several lobes compensates
for the cost of graphene bending.

The multi-lobed wrinkle structures are observed for Gr/Ir(111). Figure
\ref{fig:7}a shows a constant-current topographic STM image of a
wrinkle, indicating the existence of several lobes. In order to resolve
the structure in more detail, first derivative of the topographic
image is shown in Figure\ref{fig:7}b. Additionally, Figure \ref{fig:7}c
shows the wrinkle profile (taken along the red line in Figure \ref{fig:7}a)
where we identify four separate lobes as schematically indicated by
a dashed line. In our STM experiments such complex wrinkle profile
was often found. Possible multiple-tip effects are eliminated on the
basis of the observation that multi-lobed structures are observed
in joining wrinkles protruding in different directions (cf. Figure
\ref{fig:5}c), thus regardless of the scanning direction. Moreover,
images of surrounding flat areas do not give any indication of multiple-tip
imaging. The measured topographic wrinkle profile does not match our
model from Figure \ref{fig:6} ideally. One should take into account,
however, that our model profile is a variational \textit{ansatz} used
only to demonstrate the propensity towards formation of multi-lobed
wrinkle profiles. As such, it is not an exact solution of the problem,
and its comparison with the experiments should be performed with care.
In addition, the finite radius of the STM tip may not allow for clear
imaging of narrow trenches in between the individual lobes. Furthermore,
during STM imaging of such curved structures an increased contributions
to the tunneling current from the side of the STM tip can be expected,
which can affect the measured topographic profile. One should note,
however, that the experimentally determined lobes have similar widths,
in reasonable agreement with the model calculation which assumes the
same widths for all the lobes.

Total width of the wrinkle is 14.36 nm whereas the height is estimated
to be 0.59 nm on the basis of its average out-of-plane displacement
(cf. blue arrows in Figure \ref{fig:7}c). In Figure \ref{fig:6}b,
the red dashed line corresponds to these experimental values: $l_{1}=14.36$
nm and $A_{1}=4\cdot A_{4}=2.35$ nm. Minimum of the curve is found
at $m\approx3$, which is in rough agreement with four lobes inferred
from the STM data. At this point we would like to highlight two important
virtues of our model: (i) it predicts multi-lobed wrinkles, and (ii)
it does so in a semi-quantitative fashion, as can be seen from the
anticipated number of lobes for the wrinkle shown in Figure \ref{fig:7}.

An issue which remains to be explained is the appearance of several
lobes in a profile of a single wrinkle. Namely, lobes could be conceptually
separated from each other forming thus several single-lobed wrinkles
which would have the same energy as a multi-lobed wrinkle. However,
whereas this would indeed be the case for parallel wrinkles in essentially
1D model considered here, the inclusion of separated lobes in the
already formed, connected two-dimensional WN would require additional
energy to form new wrinkle junctions. The fact that we do observe
multi-lobed wrinkles suggests also the scenario for their formation.
As the temperature is lowered, at a transition temperature the strain
in the graphene relaxes through the formation of, likely, single-lobed
wrinkles. The transition from a flat graphene to a WN is not instantaneous
(see the movie clip in the Supplementary data), as there is disorder
in the system (it is thus difficult to speak of a well-defined critical
temperature for wrinkle formation). As the temperature is further
lowered, the existing wrinkles grow, the new ones form, and they soon
connect to each other forming the WN with many junctions. Upon further
temperature decrease, additional strain accumulates. Its relaxation
may proceed either by formation of new wrinkles, by repositioning
of the existing wrinkles or by influx of graphene in the already formed
WN, increasing thus both wrinkle amplitude and width. Although a different
network topology may be energetically preferred at lower temperatures,
its formation would require a global reconstruction of the established
WN, so that the additional strain relaxes in the already formed network.
The dynamical scenario that we propose indicates that the topological
properties of the network are fixed in some interval of elevated temperature
required to form most of the wrinkle junctions. The once formed network
frustrates the system at lower temperatures, the wrinkle junctions
behaving as rigid anchors, and the strain relaxes only by increase
of the width and amplitude of the wrinkle profiles in the network
established at elevated temperature. The influx of graphene in the
wrinkle profiles with the lowering of the temperature induces the
formation of multiple lobes. The frustration of the WN is also consistent
with the observed hysteretic behavior in the heating and cooling cycles.

Even though they were not discussed in the existing literature in
detail, wrinkles exhibiting multiple lobes were also observed in transferred
graphene grown on Cu foils \citep{Zhu2012}. In that case, wrinkles
are found in a broad range of widths. For wide wrinkles, wrinkle height
is approximately constant and relatively low ($\sim1$ nm), whereas
for narrow wrinkles it is several times larger ($\sim2-6$ nm). This
observation can be explained with the aid of our model: wide wrinkles
tend to form multi-lobed structures (wider wrinkle $\Rightarrow$
larger number of lobes) which is reflected in their constant height.
Moreover, the accompanying AFM measurements of wrinkle profiles provide
strong indication of a presence of several sequential lobes within
wide wrinkles. The concept of competition between binding and bending
energies in wrinkling phenomena is also valid for other layered van
der Waals materials. It was recently shown that the out-of-plane deviations
found in MoS\textsubscript{2} layers exhibit configurations similar
to multi-lobed wrinkles which are stable only if the total width of
such structures is sufficiently large \citep{Kushima2014}. In the
case when separation between individual lobes is reduced, they merge
and form a single-lobed wrinkle due to the lower cost of bending energy.

\section{Conclusion}

For the R0 graphene, the WN exhibits quasi-hexagonal ordering with
wrinkles extending in three preferential directions aligned with the
dense-packed atomic rows of the Ir(111) surface. Such ordering is
absent in the R30 graphene due to the reduced graphene-iridium binding.
We find that the WN can be well approximated by the VD which greatly
facilitates its description and parametrization. The formation of
wrinkles at elevated temperatures is accompanied by an increase of
LEEM reflectivity in their immediate vicinity and we propose that
the reason for this is localized, inhomogeneous stress relaxation
of the graphene lattice. Moreover, structural features of iridium
or graphene, such as wrinkles themselves or dirt particles, can act
as nucleation sites for other wrinkles. Finally, our STM results and
model calculation indicate that multi-lobed wrinkle profiles are energetically
stable and result from the frustration of the WN during sample cooldown.
Our study brings new insights into the properties of wrinkles and
WN in epitaxial graphene and is of potential use in special electronic,
optical or mechanical applications of graphene and follow-up materials.

\section*{Acknowledgments}

The financial support by the Unity Through Knowledge Fund (Grant No.
66/10) is gratefully acknowledged. Research has been carried out in
part at the Center for Functional Nanomaterials and National Synchrotron
Light Source, BNL, which are supported by the U.S. Department of Energy,
Office of Basic Energy Sciences, under Contracts No. DE-AC02-98CH10886
and DE-SC0012704. The financial support through the Center of Excellence
for Advanced Materials and Sensing Devices, research unit for Graphene
and Related 2D Structures is gratefully acknowledged.

\section*{Appendix A. Supplementary data}

Supplementary data associated with this article can be found, in the
online version, at http://dx.doi.org/10.1016/j.carbon.2015.07.059.

\section*{References}

\bibliographystyle{apsrev4-1}
\bibliography{Petrovic_wrinkles_2015}

\begin{thebibliography}{58}%
\makeatletter
\providecommand \@ifxundefined [1]{%
 \@ifx{#1\undefined}
}%
\providecommand \@ifnum [1]{%
 \ifnum #1\expandafter \@firstoftwo
 \else \expandafter \@secondoftwo
 \fi
}%
\providecommand \@ifx [1]{%
 \ifx #1\expandafter \@firstoftwo
 \else \expandafter \@secondoftwo
 \fi
}%
\providecommand \natexlab [1]{#1}%
\providecommand \enquote  [1]{``#1''}%
\providecommand \bibnamefont  [1]{#1}%
\providecommand \bibfnamefont [1]{#1}%
\providecommand \citenamefont [1]{#1}%
\providecommand \href@noop [0]{\@secondoftwo}%
\providecommand \href [0]{\begingroup \@sanitize@url \@href}%
\providecommand \@href[1]{\@@startlink{#1}\@@href}%
\providecommand \@@href[1]{\endgroup#1\@@endlink}%
\providecommand \@sanitize@url [0]{\catcode `\\12\catcode `\$12\catcode
  `\&12\catcode `\#12\catcode `\^12\catcode `\_12\catcode `\%12\relax}%
\providecommand \@@startlink[1]{}%
\providecommand \@@endlink[0]{}%
\providecommand \url  [0]{\begingroup\@sanitize@url \@url }%
\providecommand \@url [1]{\endgroup\@href {#1}{\urlprefix }}%
\providecommand \urlprefix  [0]{URL }%
\providecommand \Eprint [0]{\href }%
\providecommand \doibase [0]{http://dx.doi.org/}%
\providecommand \selectlanguage [0]{\@gobble}%
\providecommand \bibinfo  [0]{\@secondoftwo}%
\providecommand \bibfield  [0]{\@secondoftwo}%
\providecommand \translation [1]{[#1]}%
\providecommand \BibitemOpen [0]{}%
\providecommand \bibitemStop [0]{}%
\providecommand \bibitemNoStop [0]{.\EOS\space}%
\providecommand \EOS [0]{\spacefactor3000\relax}%
\providecommand \BibitemShut  [1]{\csname bibitem#1\endcsname}%
\let\auto@bib@innerbib\@empty
\bibitem [{\citenamefont {Schwierz}(2010)}]{Schwierz2010}%
  \BibitemOpen
  \bibfield  {author} {\bibinfo {author} {\bibfnamefont {F.}~\bibnamefont
  {Schwierz}},\ }\href {\doibase 10.1038/nnano.2010.89} {\bibfield  {journal}
  {\bibinfo  {journal} {Nat. Nanotechnol.}\ }\textbf {\bibinfo {volume} {5}},\
  \bibinfo {pages} {487} (\bibinfo {year} {2010})}\BibitemShut {NoStop}%
\bibitem [{\citenamefont {Bonaccorso}\ \emph {et~al.}(2010)\citenamefont
  {Bonaccorso}, \citenamefont {Sun}, \citenamefont {Hasan},\ and\ \citenamefont
  {Ferrari}}]{Bonaccorso2010}%
  \BibitemOpen
  \bibfield  {author} {\bibinfo {author} {\bibfnamefont {F.}~\bibnamefont
  {Bonaccorso}}, \bibinfo {author} {\bibfnamefont {Z.}~\bibnamefont {Sun}},
  \bibinfo {author} {\bibfnamefont {T.}~\bibnamefont {Hasan}}, \ and\ \bibinfo
  {author} {\bibfnamefont {A.~C.}\ \bibnamefont {Ferrari}},\ }\href {\doibase
  10.1038/nphoton.2010.186} {\bibfield  {journal} {\bibinfo  {journal} {Nat.
  Photonics}\ }\textbf {\bibinfo {volume} {4}},\ \bibinfo {pages} {611}
  (\bibinfo {year} {2010})}\BibitemShut {NoStop}%
\bibitem [{\citenamefont {Brownson}\ \emph {et~al.}(2011)\citenamefont
  {Brownson}, \citenamefont {Kampouris},\ and\ \citenamefont
  {Banks}}]{Brownson2011}%
  \BibitemOpen
  \bibfield  {author} {\bibinfo {author} {\bibfnamefont {D.~A.~C.}\
  \bibnamefont {Brownson}}, \bibinfo {author} {\bibfnamefont {D.~K.}\
  \bibnamefont {Kampouris}}, \ and\ \bibinfo {author} {\bibfnamefont {C.~E.}\
  \bibnamefont {Banks}},\ }\href {\doibase 10.1016/j.jpowsour.2011.02.022}
  {\bibfield  {journal} {\bibinfo  {journal} {J. Power Sources}\ }\textbf
  {\bibinfo {volume} {196}},\ \bibinfo {pages} {4873} (\bibinfo {year}
  {2011})}\BibitemShut {NoStop}%
\bibitem [{\citenamefont {Bae}\ \emph {et~al.}(2010)\citenamefont {Bae},
  \citenamefont {Kim}, \citenamefont {Lee}, \citenamefont {Xu}, \citenamefont
  {Park}, \citenamefont {Zheng} \emph {et~al.}}]{Bae2010}%
  \BibitemOpen
  \bibfield  {author} {\bibinfo {author} {\bibfnamefont {S.}~\bibnamefont
  {Bae}}, \bibinfo {author} {\bibfnamefont {H.}~\bibnamefont {Kim}}, \bibinfo
  {author} {\bibfnamefont {Y.}~\bibnamefont {Lee}}, \bibinfo {author}
  {\bibfnamefont {X.}~\bibnamefont {Xu}}, \bibinfo {author} {\bibfnamefont
  {J.-S.}\ \bibnamefont {Park}}, \bibinfo {author} {\bibfnamefont
  {Y.}~\bibnamefont {Zheng}},  \emph {et~al.},\ }\href {\doibase
  10.1038/nnano.2010.132} {\bibfield  {journal} {\bibinfo  {journal} {Nat.
  Nanotechnol.}\ }\textbf {\bibinfo {volume} {5}},\ \bibinfo {pages} {574}
  (\bibinfo {year} {2010})}\BibitemShut {NoStop}%
\bibitem [{\citenamefont {{Castro Neto}}\ \emph {et~al.}(2009)\citenamefont
  {{Castro Neto}}, \citenamefont {Guinea}, \citenamefont {Peres}, \citenamefont
  {Novoselov},\ and\ \citenamefont {Geim}}]{CastroNeto2009}%
  \BibitemOpen
  \bibfield  {author} {\bibinfo {author} {\bibfnamefont {A.~H.}\ \bibnamefont
  {{Castro Neto}}}, \bibinfo {author} {\bibfnamefont {F.}~\bibnamefont
  {Guinea}}, \bibinfo {author} {\bibfnamefont {N.~M.~R.}\ \bibnamefont
  {Peres}}, \bibinfo {author} {\bibfnamefont {K.~S.}\ \bibnamefont
  {Novoselov}}, \ and\ \bibinfo {author} {\bibfnamefont {A.~K.}\ \bibnamefont
  {Geim}},\ }\href {\doibase 10.1103/RevModPhys.81.109} {\bibfield  {journal}
  {\bibinfo  {journal} {Rev. Mod. Phys.}\ }\textbf {\bibinfo {volume} {81}},\
  \bibinfo {pages} {109} (\bibinfo {year} {2009})}\BibitemShut {NoStop}%
\bibitem [{\citenamefont {Yazyev}\ and\ \citenamefont
  {Chen}(2014)}]{Yazyev2014}%
  \BibitemOpen
  \bibfield  {author} {\bibinfo {author} {\bibfnamefont {O.~V.}\ \bibnamefont
  {Yazyev}}\ and\ \bibinfo {author} {\bibfnamefont {Y.~P.}\ \bibnamefont
  {Chen}},\ }\href {\doibase 10.1038/nnano.2014.166} {\bibfield  {journal}
  {\bibinfo  {journal} {Nat. Nanotechnol.}\ }\textbf {\bibinfo {volume} {9}},\
  \bibinfo {pages} {755} (\bibinfo {year} {2014})}\BibitemShut {NoStop}%
\bibitem [{\citenamefont {Sutter}\ \emph {et~al.}(2009)\citenamefont {Sutter},
  \citenamefont {Sadowski},\ and\ \citenamefont {Sutter}}]{Sutter2009a}%
  \BibitemOpen
  \bibfield  {author} {\bibinfo {author} {\bibfnamefont {P.}~\bibnamefont
  {Sutter}}, \bibinfo {author} {\bibfnamefont {J.~T.}\ \bibnamefont
  {Sadowski}}, \ and\ \bibinfo {author} {\bibfnamefont {E.}~\bibnamefont
  {Sutter}},\ }\href {\doibase 10.1103/PhysRevB.80.245411} {\bibfield
  {journal} {\bibinfo  {journal} {Phys. Rev. B}\ }\textbf {\bibinfo {volume}
  {80}},\ \bibinfo {pages} {245411} (\bibinfo {year} {2009})}\BibitemShut
  {NoStop}%
\bibitem [{\citenamefont {Chae}\ \emph {et~al.}(2009)\citenamefont {Chae},
  \citenamefont {G\"{u}neş}, \citenamefont {Kim}, \citenamefont {Kim},
  \citenamefont {Han}, \citenamefont {Kim} \emph {et~al.}}]{Chae2009}%
  \BibitemOpen
  \bibfield  {author} {\bibinfo {author} {\bibfnamefont {S.~J.}\ \bibnamefont
  {Chae}}, \bibinfo {author} {\bibfnamefont {F.}~\bibnamefont {G\"{u}neş}},
  \bibinfo {author} {\bibfnamefont {K.~K.}\ \bibnamefont {Kim}}, \bibinfo
  {author} {\bibfnamefont {E.~S.}\ \bibnamefont {Kim}}, \bibinfo {author}
  {\bibfnamefont {G.~H.}\ \bibnamefont {Han}}, \bibinfo {author} {\bibfnamefont
  {S.~M.}\ \bibnamefont {Kim}},  \emph {et~al.},\ }\href {\doibase
  10.1002/adma.200803016} {\bibfield  {journal} {\bibinfo  {journal} {Adv.
  Mater.}\ }\textbf {\bibinfo {volume} {21}},\ \bibinfo {pages} {2328}
  (\bibinfo {year} {2009})}\BibitemShut {NoStop}%
\bibitem [{\citenamefont {Obraztsov}\ \emph {et~al.}(2007)\citenamefont
  {Obraztsov}, \citenamefont {Obraztsova}, \citenamefont {Tyurnina},\ and\
  \citenamefont {Zolotukhin}}]{Obraztsov2007}%
  \BibitemOpen
  \bibfield  {author} {\bibinfo {author} {\bibfnamefont {A.~N.}\ \bibnamefont
  {Obraztsov}}, \bibinfo {author} {\bibfnamefont {E.~A.}\ \bibnamefont
  {Obraztsova}}, \bibinfo {author} {\bibfnamefont {A.~V.}\ \bibnamefont
  {Tyurnina}}, \ and\ \bibinfo {author} {\bibfnamefont {A.~A.}\ \bibnamefont
  {Zolotukhin}},\ }\href {\doibase 10.1016/j.carbon.2007.05.028} {\bibfield
  {journal} {\bibinfo  {journal} {Carbon}\ }\textbf {\bibinfo {volume} {45}},\
  \bibinfo {pages} {2017} (\bibinfo {year} {2007})}\BibitemShut {NoStop}%
\bibitem [{\citenamefont {N'Diaye}\ \emph {et~al.}(2009)\citenamefont
  {N'Diaye}, \citenamefont {{Van Gastel}}, \citenamefont
  {Mart\'{\i}nez-Galera}, \citenamefont {Coraux}, \citenamefont {Hattab},
  \citenamefont {Wall} \emph {et~al.}}]{N'Diaye2009b}%
  \BibitemOpen
  \bibfield  {author} {\bibinfo {author} {\bibfnamefont {A.~T.}\ \bibnamefont
  {N'Diaye}}, \bibinfo {author} {\bibfnamefont {R.}~\bibnamefont {{Van
  Gastel}}}, \bibinfo {author} {\bibfnamefont {A.~J.}\ \bibnamefont
  {Mart\'{\i}nez-Galera}}, \bibinfo {author} {\bibfnamefont {J.}~\bibnamefont
  {Coraux}}, \bibinfo {author} {\bibfnamefont {H.}~\bibnamefont {Hattab}},
  \bibinfo {author} {\bibfnamefont {D.}~\bibnamefont {Wall}},  \emph {et~al.},\
  }\href {\doibase 10.1088/1367-2630/11/11/113056} {\bibfield  {journal}
  {\bibinfo  {journal} {New J. Phys.}\ }\textbf {\bibinfo {volume} {11}},\
  \bibinfo {pages} {113056} (\bibinfo {year} {2009})}\BibitemShut {NoStop}%
\bibitem [{\citenamefont {Loginova}\ \emph {et~al.}(2009)\citenamefont
  {Loginova}, \citenamefont {Nie}, \citenamefont {Th\"{u}rmer}, \citenamefont
  {Bartelt},\ and\ \citenamefont {McCarty}}]{Loginova2009a}%
  \BibitemOpen
  \bibfield  {author} {\bibinfo {author} {\bibfnamefont {E.}~\bibnamefont
  {Loginova}}, \bibinfo {author} {\bibfnamefont {S.}~\bibnamefont {Nie}},
  \bibinfo {author} {\bibfnamefont {K.}~\bibnamefont {Th\"{u}rmer}}, \bibinfo
  {author} {\bibfnamefont {N.~C.}\ \bibnamefont {Bartelt}}, \ and\ \bibinfo
  {author} {\bibfnamefont {K.~F.}\ \bibnamefont {McCarty}},\ }\href {\doibase
  10.1103/PhysRevB.80.085430} {\bibfield  {journal} {\bibinfo  {journal} {Phys.
  Rev. B}\ }\textbf {\bibinfo {volume} {80}},\ \bibinfo {pages} {085430}
  (\bibinfo {year} {2009})}\BibitemShut {NoStop}%
\bibitem [{\citenamefont {Li}\ \emph {et~al.}(2009)\citenamefont {Li},
  \citenamefont {Cai}, \citenamefont {An}, \citenamefont {Kim}, \citenamefont
  {Nah}, \citenamefont {Yang} \emph {et~al.}}]{Li2009}%
  \BibitemOpen
  \bibfield  {author} {\bibinfo {author} {\bibfnamefont {X.}~\bibnamefont
  {Li}}, \bibinfo {author} {\bibfnamefont {W.}~\bibnamefont {Cai}}, \bibinfo
  {author} {\bibfnamefont {J.}~\bibnamefont {An}}, \bibinfo {author}
  {\bibfnamefont {S.}~\bibnamefont {Kim}}, \bibinfo {author} {\bibfnamefont
  {J.}~\bibnamefont {Nah}}, \bibinfo {author} {\bibfnamefont {D.}~\bibnamefont
  {Yang}},  \emph {et~al.},\ }\href {\doibase 10.1126/science.1171245}
  {\bibfield  {journal} {\bibinfo  {journal} {Science}\ }\textbf {\bibinfo
  {volume} {324}},\ \bibinfo {pages} {1312} (\bibinfo {year}
  {2009})}\BibitemShut {NoStop}%
\bibitem [{\citenamefont {Liu}\ \emph {et~al.}(2012)\citenamefont {Liu},
  \citenamefont {Zhang}, \citenamefont {Chen}, \citenamefont {Gao},
  \citenamefont {Gao}, \citenamefont {Ma} \emph {et~al.}}]{Liu2012a}%
  \BibitemOpen
  \bibfield  {author} {\bibinfo {author} {\bibfnamefont {M.}~\bibnamefont
  {Liu}}, \bibinfo {author} {\bibfnamefont {Y.}~\bibnamefont {Zhang}}, \bibinfo
  {author} {\bibfnamefont {Y.}~\bibnamefont {Chen}}, \bibinfo {author}
  {\bibfnamefont {Y.}~\bibnamefont {Gao}}, \bibinfo {author} {\bibfnamefont
  {T.}~\bibnamefont {Gao}}, \bibinfo {author} {\bibfnamefont {D.}~\bibnamefont
  {Ma}},  \emph {et~al.},\ }\href {\doibase 10.1021/nn3047154} {\bibfield
  {journal} {\bibinfo  {journal} {ACS Nano}\ }\textbf {\bibinfo {volume} {6}},\
  \bibinfo {pages} {10581} (\bibinfo {year} {2012})}\BibitemShut {NoStop}%
\bibitem [{\citenamefont {Biedermann}\ \emph {et~al.}(2009)\citenamefont
  {Biedermann}, \citenamefont {Bolen}, \citenamefont {Capano}, \citenamefont
  {Zemlyanov},\ and\ \citenamefont {Reifenberger}}]{Biedermann2009}%
  \BibitemOpen
  \bibfield  {author} {\bibinfo {author} {\bibfnamefont {L.~B.}\ \bibnamefont
  {Biedermann}}, \bibinfo {author} {\bibfnamefont {M.~L.}\ \bibnamefont
  {Bolen}}, \bibinfo {author} {\bibfnamefont {M.~A.}\ \bibnamefont {Capano}},
  \bibinfo {author} {\bibfnamefont {D.}~\bibnamefont {Zemlyanov}}, \ and\
  \bibinfo {author} {\bibfnamefont {R.~G.}\ \bibnamefont {Reifenberger}},\
  }\href {\doibase 10.1103/PhysRevB.79.125411} {\bibfield  {journal} {\bibinfo
  {journal} {Phys. Rev. B}\ }\textbf {\bibinfo {volume} {79}},\ \bibinfo
  {pages} {125411} (\bibinfo {year} {2009})}\BibitemShut {NoStop}%
\bibitem [{\citenamefont {Sun}\ \emph {et~al.}(2009)\citenamefont {Sun},
  \citenamefont {Jia}, \citenamefont {Xue},\ and\ \citenamefont
  {Li}}]{Sun2009}%
  \BibitemOpen
  \bibfield  {author} {\bibinfo {author} {\bibfnamefont {G.~F.}\ \bibnamefont
  {Sun}}, \bibinfo {author} {\bibfnamefont {J.~F.}\ \bibnamefont {Jia}},
  \bibinfo {author} {\bibfnamefont {Q.~K.}\ \bibnamefont {Xue}}, \ and\
  \bibinfo {author} {\bibfnamefont {L.}~\bibnamefont {Li}},\ }\href {\doibase
  10.1088/0957-4484/20/35/355701} {\bibfield  {journal} {\bibinfo  {journal}
  {Nanotechnology}\ }\textbf {\bibinfo {volume} {20}},\ \bibinfo {pages}
  {355701} (\bibinfo {year} {2009})}\BibitemShut {NoStop}%
\bibitem [{\citenamefont {Prakash}\ \emph {et~al.}(2010)\citenamefont
  {Prakash}, \citenamefont {Capano}, \citenamefont {Bolen}, \citenamefont
  {Zemlyanov},\ and\ \citenamefont {Reifenberger}}]{Prakash2010}%
  \BibitemOpen
  \bibfield  {author} {\bibinfo {author} {\bibfnamefont {G.}~\bibnamefont
  {Prakash}}, \bibinfo {author} {\bibfnamefont {M.~A.}\ \bibnamefont {Capano}},
  \bibinfo {author} {\bibfnamefont {M.~L.}\ \bibnamefont {Bolen}}, \bibinfo
  {author} {\bibfnamefont {D.}~\bibnamefont {Zemlyanov}}, \ and\ \bibinfo
  {author} {\bibfnamefont {R.~G.}\ \bibnamefont {Reifenberger}},\ }\href
  {\doibase 10.1016/j.carbon.2010.02.026} {\bibfield  {journal} {\bibinfo
  {journal} {Carbon}\ }\textbf {\bibinfo {volume} {48}},\ \bibinfo {pages}
  {2383} (\bibinfo {year} {2010})}\BibitemShut {NoStop}%
\bibitem [{\citenamefont {Lanza}\ \emph {et~al.}(2013)\citenamefont {Lanza},
  \citenamefont {Wang}, \citenamefont {Bayerl}, \citenamefont {Gao},
  \citenamefont {Porti}, \citenamefont {Nafria} \emph {et~al.}}]{Lanza2013b}%
  \BibitemOpen
  \bibfield  {author} {\bibinfo {author} {\bibfnamefont {M.}~\bibnamefont
  {Lanza}}, \bibinfo {author} {\bibfnamefont {Y.}~\bibnamefont {Wang}},
  \bibinfo {author} {\bibfnamefont {A.}~\bibnamefont {Bayerl}}, \bibinfo
  {author} {\bibfnamefont {T.}~\bibnamefont {Gao}}, \bibinfo {author}
  {\bibfnamefont {M.}~\bibnamefont {Porti}}, \bibinfo {author} {\bibfnamefont
  {M.}~\bibnamefont {Nafria}},  \emph {et~al.},\ }\href {\doibase
  10.1063/1.4794521} {\bibfield  {journal} {\bibinfo  {journal} {J. Appl.
  Phys.}\ }\textbf {\bibinfo {volume} {113}},\ \bibinfo {pages} {104301}
  (\bibinfo {year} {2013})}\BibitemShut {NoStop}%
\bibitem [{\citenamefont {Zhu}\ \emph {et~al.}(2012)\citenamefont {Zhu},
  \citenamefont {Low}, \citenamefont {Perebeinos}, \citenamefont {Bol},
  \citenamefont {Zhu}, \citenamefont {Yan} \emph {et~al.}}]{Zhu2012}%
  \BibitemOpen
  \bibfield  {author} {\bibinfo {author} {\bibfnamefont {W.}~\bibnamefont
  {Zhu}}, \bibinfo {author} {\bibfnamefont {T.}~\bibnamefont {Low}}, \bibinfo
  {author} {\bibfnamefont {V.}~\bibnamefont {Perebeinos}}, \bibinfo {author}
  {\bibfnamefont {A.~A.}\ \bibnamefont {Bol}}, \bibinfo {author} {\bibfnamefont
  {Y.}~\bibnamefont {Zhu}}, \bibinfo {author} {\bibfnamefont {H.}~\bibnamefont
  {Yan}},  \emph {et~al.},\ }\href {\doibase 10.1021/nl300563h} {\bibfield
  {journal} {\bibinfo  {journal} {Nano Lett.}\ }\textbf {\bibinfo {volume}
  {12}},\ \bibinfo {pages} {3431} (\bibinfo {year} {2012})}\BibitemShut
  {NoStop}%
\bibitem [{\citenamefont {Xu}\ \emph {et~al.}(2009)\citenamefont {Xu},
  \citenamefont {Cao},\ and\ \citenamefont {Heath}}]{Xu2009}%
  \BibitemOpen
  \bibfield  {author} {\bibinfo {author} {\bibfnamefont {K.}~\bibnamefont
  {Xu}}, \bibinfo {author} {\bibfnamefont {P.}~\bibnamefont {Cao}}, \ and\
  \bibinfo {author} {\bibfnamefont {J.~R.}\ \bibnamefont {Heath}},\ }\href
  {\doibase 10.1021/nl902729p} {\bibfield  {journal} {\bibinfo  {journal} {Nano
  Lett.}\ }\textbf {\bibinfo {volume} {9}},\ \bibinfo {pages} {4446} (\bibinfo
  {year} {2009})}\BibitemShut {NoStop}%
\bibitem [{\citenamefont {Gao}\ \emph {et~al.}(2012)\citenamefont {Gao},
  \citenamefont {Ren}, \citenamefont {Xu}, \citenamefont {Jin}, \citenamefont
  {Wang}, \citenamefont {Ma} \emph {et~al.}}]{Gao2012}%
  \BibitemOpen
  \bibfield  {author} {\bibinfo {author} {\bibfnamefont {L.}~\bibnamefont
  {Gao}}, \bibinfo {author} {\bibfnamefont {W.}~\bibnamefont {Ren}}, \bibinfo
  {author} {\bibfnamefont {H.}~\bibnamefont {Xu}}, \bibinfo {author}
  {\bibfnamefont {L.}~\bibnamefont {Jin}}, \bibinfo {author} {\bibfnamefont
  {Z.}~\bibnamefont {Wang}}, \bibinfo {author} {\bibfnamefont {T.}~\bibnamefont
  {Ma}},  \emph {et~al.},\ }\href {\doibase 10.1038/ncomms1702} {\bibfield
  {journal} {\bibinfo  {journal} {Nat. Commun.}\ }\textbf {\bibinfo {volume}
  {3}},\ \bibinfo {pages} {699} (\bibinfo {year} {2012})}\BibitemShut {NoStop}%
\bibitem [{\citenamefont {Kim}\ \emph {et~al.}(2012)\citenamefont {Kim},
  \citenamefont {Hsu}, \citenamefont {Jia}, \citenamefont {Kim}, \citenamefont
  {Shi}, \citenamefont {Hofmann} \emph {et~al.}}]{Kim2012}%
  \BibitemOpen
  \bibfield  {author} {\bibinfo {author} {\bibfnamefont {K.~K.}\ \bibnamefont
  {Kim}}, \bibinfo {author} {\bibfnamefont {A.}~\bibnamefont {Hsu}}, \bibinfo
  {author} {\bibfnamefont {X.}~\bibnamefont {Jia}}, \bibinfo {author}
  {\bibfnamefont {S.~M.}\ \bibnamefont {Kim}}, \bibinfo {author} {\bibfnamefont
  {Y.}~\bibnamefont {Shi}}, \bibinfo {author} {\bibfnamefont {M.}~\bibnamefont
  {Hofmann}},  \emph {et~al.},\ }\href {\doibase 10.1021/nl203249a} {\bibfield
  {journal} {\bibinfo  {journal} {Nano Lett.}\ }\textbf {\bibinfo {volume}
  {12}},\ \bibinfo {pages} {161} (\bibinfo {year} {2012})}\BibitemShut
  {NoStop}%
\bibitem [{\citenamefont {Kushima}\ \emph {et~al.}(2015)\citenamefont
  {Kushima}, \citenamefont {Qian}, \citenamefont {Zhao}, \citenamefont
  {Zhang},\ and\ \citenamefont {Li}}]{Kushima2014}%
  \BibitemOpen
  \bibfield  {author} {\bibinfo {author} {\bibfnamefont {A.}~\bibnamefont
  {Kushima}}, \bibinfo {author} {\bibfnamefont {X.}~\bibnamefont {Qian}},
  \bibinfo {author} {\bibfnamefont {P.}~\bibnamefont {Zhao}}, \bibinfo {author}
  {\bibfnamefont {S.}~\bibnamefont {Zhang}}, \ and\ \bibinfo {author}
  {\bibfnamefont {J.}~\bibnamefont {Li}},\ }\href {\doibase 10.1021/nl5045082}
  {\bibfield  {journal} {\bibinfo  {journal} {Nano Lett.}\ }\textbf {\bibinfo
  {volume} {15}},\ \bibinfo {pages} {1302} (\bibinfo {year}
  {2015})}\BibitemShut {NoStop}%
\bibitem [{\citenamefont {Mei}\ \emph {et~al.}(2007)\citenamefont {Mei},
  \citenamefont {Thurmer}, \citenamefont {Cavallo}, \citenamefont
  {Kiravittaya},\ and\ \citenamefont {Schmidt}}]{Mei2007}%
  \BibitemOpen
  \bibfield  {author} {\bibinfo {author} {\bibfnamefont {Y.}~\bibnamefont
  {Mei}}, \bibinfo {author} {\bibfnamefont {D.~J.}\ \bibnamefont {Thurmer}},
  \bibinfo {author} {\bibfnamefont {F.}~\bibnamefont {Cavallo}}, \bibinfo
  {author} {\bibfnamefont {S.}~\bibnamefont {Kiravittaya}}, \ and\ \bibinfo
  {author} {\bibfnamefont {O.~G.}\ \bibnamefont {Schmidt}},\ }\href {\doibase
  10.1002/adma.200601622} {\bibfield  {journal} {\bibinfo  {journal} {Adv.
  Mater.}\ }\textbf {\bibinfo {volume} {19}},\ \bibinfo {pages} {2124}
  (\bibinfo {year} {2007})}\BibitemShut {NoStop}%
\bibitem [{\citenamefont {Li}\ \emph {et~al.}(2012)\citenamefont {Li},
  \citenamefont {Cao}, \citenamefont {Feng},\ and\ \citenamefont
  {Gao}}]{Li2012}%
  \BibitemOpen
  \bibfield  {author} {\bibinfo {author} {\bibfnamefont {B.}~\bibnamefont
  {Li}}, \bibinfo {author} {\bibfnamefont {Y.-P.}\ \bibnamefont {Cao}},
  \bibinfo {author} {\bibfnamefont {X.-Q.}\ \bibnamefont {Feng}}, \ and\
  \bibinfo {author} {\bibfnamefont {H.}~\bibnamefont {Gao}},\ }\href {\doibase
  10.1039/c2sm00011c} {\bibfield  {journal} {\bibinfo  {journal} {Soft Matter}\
  }\textbf {\bibinfo {volume} {8}},\ \bibinfo {pages} {5728} (\bibinfo {year}
  {2012})}\BibitemShut {NoStop}%
\bibitem [{\citenamefont {Yazyev}\ and\ \citenamefont
  {Louie}(2010)}]{Yazyev2010}%
  \BibitemOpen
  \bibfield  {author} {\bibinfo {author} {\bibfnamefont {O.~V.}\ \bibnamefont
  {Yazyev}}\ and\ \bibinfo {author} {\bibfnamefont {S.~G.}\ \bibnamefont
  {Louie}},\ }\href {\doibase 10.1038/nmat2830} {\bibfield  {journal} {\bibinfo
   {journal} {Nat. Mater.}\ }\textbf {\bibinfo {volume} {9}},\ \bibinfo {pages}
  {806} (\bibinfo {year} {2010})}\BibitemShut {NoStop}%
\bibitem [{\citenamefont {Chen}\ \emph {et~al.}(2012)\citenamefont {Chen},
  \citenamefont {Li}, \citenamefont {Zhang}, \citenamefont {Qu}, \citenamefont
  {Ji}, \citenamefont {Ruoff} \emph {et~al.}}]{Chen2012}%
  \BibitemOpen
  \bibfield  {author} {\bibinfo {author} {\bibfnamefont {S.}~\bibnamefont
  {Chen}}, \bibinfo {author} {\bibfnamefont {Q.}~\bibnamefont {Li}}, \bibinfo
  {author} {\bibfnamefont {Q.}~\bibnamefont {Zhang}}, \bibinfo {author}
  {\bibfnamefont {Y.}~\bibnamefont {Qu}}, \bibinfo {author} {\bibfnamefont
  {H.}~\bibnamefont {Ji}}, \bibinfo {author} {\bibfnamefont {R.~S.}\
  \bibnamefont {Ruoff}},  \emph {et~al.},\ }\href {\doibase
  10.1088/0957-4484/23/36/365701} {\bibfield  {journal} {\bibinfo  {journal}
  {Nanotechnology}\ }\textbf {\bibinfo {volume} {23}},\ \bibinfo {pages}
  {365701} (\bibinfo {year} {2012})}\BibitemShut {NoStop}%
\bibitem [{\citenamefont {Srivastava}\ \emph {et~al.}(1999)\citenamefont
  {Srivastava}, \citenamefont {Brenner}, \citenamefont {Schall}, \citenamefont
  {Ausman}, \citenamefont {Yu},\ and\ \citenamefont {Ruoff}}]{Srivastava1999}%
  \BibitemOpen
  \bibfield  {author} {\bibinfo {author} {\bibfnamefont {D.}~\bibnamefont
  {Srivastava}}, \bibinfo {author} {\bibfnamefont {D.~W.}\ \bibnamefont
  {Brenner}}, \bibinfo {author} {\bibfnamefont {J.~D.}\ \bibnamefont {Schall}},
  \bibinfo {author} {\bibfnamefont {K.~D.}\ \bibnamefont {Ausman}}, \bibinfo
  {author} {\bibfnamefont {M.}~\bibnamefont {Yu}}, \ and\ \bibinfo {author}
  {\bibfnamefont {R.~S.}\ \bibnamefont {Ruoff}},\ }\href {\doibase
  10.1021/jp990882s} {\bibfield  {journal} {\bibinfo  {journal} {J. Phys. Chem.
  B}\ }\textbf {\bibinfo {volume} {103}},\ \bibinfo {pages} {4330} (\bibinfo
  {year} {1999})}\BibitemShut {NoStop}%
\bibitem [{\citenamefont {Starodub}\ \emph {et~al.}(2010)\citenamefont
  {Starodub}, \citenamefont {Bartelt},\ and\ \citenamefont
  {McCarty}}]{Starodub2010a}%
  \BibitemOpen
  \bibfield  {author} {\bibinfo {author} {\bibfnamefont {E.}~\bibnamefont
  {Starodub}}, \bibinfo {author} {\bibfnamefont {N.~C.}\ \bibnamefont
  {Bartelt}}, \ and\ \bibinfo {author} {\bibfnamefont {K.~F.}\ \bibnamefont
  {McCarty}},\ }\href {\doibase 10.1021/jp912139e} {\bibfield  {journal}
  {\bibinfo  {journal} {J. Phys. Chem. C}\ }\textbf {\bibinfo {volume} {114}},\
  \bibinfo {pages} {5134} (\bibinfo {year} {2010})}\BibitemShut {NoStop}%
\bibitem [{\citenamefont {Zhang}\ \emph {et~al.}(2013)\citenamefont {Zhang},
  \citenamefont {Fu}, \citenamefont {Cui}, \citenamefont {Mu}, \citenamefont
  {Jin},\ and\ \citenamefont {Bao}}]{Zhang2013PCCP}%
  \BibitemOpen
  \bibfield  {author} {\bibinfo {author} {\bibfnamefont {Y.}~\bibnamefont
  {Zhang}}, \bibinfo {author} {\bibfnamefont {Q.}~\bibnamefont {Fu}}, \bibinfo
  {author} {\bibfnamefont {Y.}~\bibnamefont {Cui}}, \bibinfo {author}
  {\bibfnamefont {R.}~\bibnamefont {Mu}}, \bibinfo {author} {\bibfnamefont
  {L.}~\bibnamefont {Jin}}, \ and\ \bibinfo {author} {\bibfnamefont
  {X.}~\bibnamefont {Bao}},\ }\href {\doibase 10.1039/c3cp52115j} {\bibfield
  {journal} {\bibinfo  {journal} {Phys. Chem. Chem. Phys.}\ }\textbf {\bibinfo
  {volume} {15}},\ \bibinfo {pages} {19042} (\bibinfo {year}
  {2013})}\BibitemShut {NoStop}%
\bibitem [{\citenamefont {Khokhar}\ \emph {et~al.}(2012)\citenamefont
  {Khokhar}, \citenamefont {Hlawacek}, \citenamefont {{Van Gastel}},
  \citenamefont {Zandvliet}, \citenamefont {Teichert},\ and\ \citenamefont
  {Poelsema}}]{Khokhar2012}%
  \BibitemOpen
  \bibfield  {author} {\bibinfo {author} {\bibfnamefont {F.~S.}\ \bibnamefont
  {Khokhar}}, \bibinfo {author} {\bibfnamefont {G.}~\bibnamefont {Hlawacek}},
  \bibinfo {author} {\bibfnamefont {R.}~\bibnamefont {{Van Gastel}}}, \bibinfo
  {author} {\bibfnamefont {H.~J.~W.}\ \bibnamefont {Zandvliet}}, \bibinfo
  {author} {\bibfnamefont {C.}~\bibnamefont {Teichert}}, \ and\ \bibinfo
  {author} {\bibfnamefont {B.}~\bibnamefont {Poelsema}},\ }\href {\doibase
  10.1016/j.susc.2011.11.012} {\bibfield  {journal} {\bibinfo  {journal} {Surf.
  Sci.}\ }\textbf {\bibinfo {volume} {606}},\ \bibinfo {pages} {475} (\bibinfo
  {year} {2012})}\BibitemShut {NoStop}%
\bibitem [{\citenamefont {Petrovi\'{c}}\ \emph {et~al.}(2013)\citenamefont
  {Petrovi\'{c}}, \citenamefont {{\v{S}rut Raki\'{c}}}, \citenamefont {Runte},
  \citenamefont {Busse}, \citenamefont {Sadowski}, \citenamefont {Lazi\'{c}}
  \emph {et~al.}}]{Petrovic2013a}%
  \BibitemOpen
  \bibfield  {author} {\bibinfo {author} {\bibfnamefont {M.}~\bibnamefont
  {Petrovi\'{c}}}, \bibinfo {author} {\bibfnamefont {I.}~\bibnamefont
  {{\v{S}rut Raki\'{c}}}}, \bibinfo {author} {\bibfnamefont {S.}~\bibnamefont
  {Runte}}, \bibinfo {author} {\bibfnamefont {C.}~\bibnamefont {Busse}},
  \bibinfo {author} {\bibfnamefont {J.~T.}\ \bibnamefont {Sadowski}}, \bibinfo
  {author} {\bibfnamefont {P.}~\bibnamefont {Lazi\'{c}}},  \emph {et~al.},\
  }\href {\doibase 10.1038/ncomms3772} {\bibfield  {journal} {\bibinfo
  {journal} {Nat. Commun.}\ }\textbf {\bibinfo {volume} {4}},\ \bibinfo {pages}
  {2772} (\bibinfo {year} {2013})}\BibitemShut {NoStop}%
\bibitem [{\citenamefont {Schumacher}\ \emph {et~al.}(2014)\citenamefont
  {Schumacher}, \citenamefont {Huttmann}, \citenamefont {Petrovi\'{c}},
  \citenamefont {Witt}, \citenamefont {F\"{o}rster}, \citenamefont {Vo-Van}
  \emph {et~al.}}]{Schumacher2014}%
  \BibitemOpen
  \bibfield  {author} {\bibinfo {author} {\bibfnamefont {S.}~\bibnamefont
  {Schumacher}}, \bibinfo {author} {\bibfnamefont {F.}~\bibnamefont
  {Huttmann}}, \bibinfo {author} {\bibfnamefont {M.}~\bibnamefont
  {Petrovi\'{c}}}, \bibinfo {author} {\bibfnamefont {C.}~\bibnamefont {Witt}},
  \bibinfo {author} {\bibfnamefont {D.~F.}\ \bibnamefont {F\"{o}rster}},
  \bibinfo {author} {\bibfnamefont {C.}~\bibnamefont {Vo-Van}},  \emph
  {et~al.},\ }\href {\doibase 10.1103/PhysRevB.90.235437} {\bibfield  {journal}
  {\bibinfo  {journal} {Phys. Rev. B}\ }\textbf {\bibinfo {volume} {90}},\
  \bibinfo {pages} {235437} (\bibinfo {year} {2014})}\BibitemShut {NoStop}%
\bibitem [{\citenamefont {Vlaic}\ \emph {et~al.}(2014)\citenamefont {Vlaic},
  \citenamefont {Kimouche}, \citenamefont {Coraux}, \citenamefont {Santos},
  \citenamefont {Locatelli},\ and\ \citenamefont {Rougemaille}}]{Vlaic2014}%
  \BibitemOpen
  \bibfield  {author} {\bibinfo {author} {\bibfnamefont {S.}~\bibnamefont
  {Vlaic}}, \bibinfo {author} {\bibfnamefont {A.}~\bibnamefont {Kimouche}},
  \bibinfo {author} {\bibfnamefont {J.}~\bibnamefont {Coraux}}, \bibinfo
  {author} {\bibfnamefont {B.}~\bibnamefont {Santos}}, \bibinfo {author}
  {\bibfnamefont {A.}~\bibnamefont {Locatelli}}, \ and\ \bibinfo {author}
  {\bibfnamefont {N.}~\bibnamefont {Rougemaille}},\ }\href {\doibase
  10.1063/1.4868119} {\bibfield  {journal} {\bibinfo  {journal} {Appl. Phys.
  Lett.}\ }\textbf {\bibinfo {volume} {104}},\ \bibinfo {pages} {101602}
  (\bibinfo {year} {2014})}\BibitemShut {NoStop}%
\bibitem [{\citenamefont {Kimouche}\ \emph {et~al.}(2014)\citenamefont
  {Kimouche}, \citenamefont {Renault}, \citenamefont {Samaddar}, \citenamefont
  {Winkelmann}, \citenamefont {Courtois}, \citenamefont {Fruchart} \emph
  {et~al.}}]{Kimouche2014}%
  \BibitemOpen
  \bibfield  {author} {\bibinfo {author} {\bibfnamefont {A.}~\bibnamefont
  {Kimouche}}, \bibinfo {author} {\bibfnamefont {O.}~\bibnamefont {Renault}},
  \bibinfo {author} {\bibfnamefont {S.}~\bibnamefont {Samaddar}}, \bibinfo
  {author} {\bibfnamefont {C.}~\bibnamefont {Winkelmann}}, \bibinfo {author}
  {\bibfnamefont {H.}~\bibnamefont {Courtois}}, \bibinfo {author}
  {\bibfnamefont {O.}~\bibnamefont {Fruchart}},  \emph {et~al.},\ }\href
  {\doibase 10.1016/j.carbon.2013.10.033} {\bibfield  {journal} {\bibinfo
  {journal} {Carbon}\ }\textbf {\bibinfo {volume} {68}},\ \bibinfo {pages} {73}
  (\bibinfo {year} {2014})}\BibitemShut {NoStop}%
\bibitem [{\citenamefont {Zang}\ \emph {et~al.}(2013)\citenamefont {Zang},
  \citenamefont {Ryu}, \citenamefont {Pugno}, \citenamefont {Wang},
  \citenamefont {Tu}, \citenamefont {Buehler} \emph {et~al.}}]{Zang2013}%
  \BibitemOpen
  \bibfield  {author} {\bibinfo {author} {\bibfnamefont {J.}~\bibnamefont
  {Zang}}, \bibinfo {author} {\bibfnamefont {S.}~\bibnamefont {Ryu}}, \bibinfo
  {author} {\bibfnamefont {N.}~\bibnamefont {Pugno}}, \bibinfo {author}
  {\bibfnamefont {Q.}~\bibnamefont {Wang}}, \bibinfo {author} {\bibfnamefont
  {Q.}~\bibnamefont {Tu}}, \bibinfo {author} {\bibfnamefont {M.~J.}\
  \bibnamefont {Buehler}},  \emph {et~al.},\ }\href {\doibase 10.1038/nmat3542}
  {\bibfield  {journal} {\bibinfo  {journal} {Nat. Mater.}\ }\textbf {\bibinfo
  {volume} {12}},\ \bibinfo {pages} {321} (\bibinfo {year} {2013})}\BibitemShut
  {NoStop}%
\bibitem [{\citenamefont {Pan}\ \emph {et~al.}(2011)\citenamefont {Pan},
  \citenamefont {Liu}, \citenamefont {Fu},\ and\ \citenamefont
  {Liu}}]{Pan2011b}%
  \BibitemOpen
  \bibfield  {author} {\bibinfo {author} {\bibfnamefont {Z.}~\bibnamefont
  {Pan}}, \bibinfo {author} {\bibfnamefont {N.}~\bibnamefont {Liu}}, \bibinfo
  {author} {\bibfnamefont {L.}~\bibnamefont {Fu}}, \ and\ \bibinfo {author}
  {\bibfnamefont {Z.}~\bibnamefont {Liu}},\ }\href {\doibase 10.1021/ja207517u}
  {\bibfield  {journal} {\bibinfo  {journal} {J. Am. Chem. Soc.}\ }\textbf
  {\bibinfo {volume} {133}},\ \bibinfo {pages} {17578} (\bibinfo {year}
  {2011})}\BibitemShut {NoStop}%
\bibitem [{\citenamefont {Kim}\ \emph {et~al.}(2011)\citenamefont {Kim},
  \citenamefont {Lee}, \citenamefont {Malone}, \citenamefont {Chan},
  \citenamefont {Alem\'{a}n}, \citenamefont {Regan} \emph {et~al.}}]{Kim2011a}%
  \BibitemOpen
  \bibfield  {author} {\bibinfo {author} {\bibfnamefont {K.}~\bibnamefont
  {Kim}}, \bibinfo {author} {\bibfnamefont {Z.}~\bibnamefont {Lee}}, \bibinfo
  {author} {\bibfnamefont {B.~D.}\ \bibnamefont {Malone}}, \bibinfo {author}
  {\bibfnamefont {K.~T.}\ \bibnamefont {Chan}}, \bibinfo {author}
  {\bibfnamefont {B.}~\bibnamefont {Alem\'{a}n}}, \bibinfo {author}
  {\bibfnamefont {W.}~\bibnamefont {Regan}},  \emph {et~al.},\ }\href {\doibase
  10.1103/PhysRevB.83.245433} {\bibfield  {journal} {\bibinfo  {journal} {Phys.
  Rev. B}\ }\textbf {\bibinfo {volume} {83}},\ \bibinfo {pages} {245433}
  (\bibinfo {year} {2011})}\BibitemShut {NoStop}%
\bibitem [{\citenamefont {Coraux}\ \emph {et~al.}(2008)\citenamefont {Coraux},
  \citenamefont {N'Diaye}, \citenamefont {Busse},\ and\ \citenamefont
  {Michely}}]{Coraux2008}%
  \BibitemOpen
  \bibfield  {author} {\bibinfo {author} {\bibfnamefont {J.}~\bibnamefont
  {Coraux}}, \bibinfo {author} {\bibfnamefont {A.~T.}\ \bibnamefont {N'Diaye}},
  \bibinfo {author} {\bibfnamefont {C.}~\bibnamefont {Busse}}, \ and\ \bibinfo
  {author} {\bibfnamefont {T.}~\bibnamefont {Michely}},\ }\href {\doibase
  10.1021/nl0728874} {\bibfield  {journal} {\bibinfo  {journal} {Nano Lett.}\
  }\textbf {\bibinfo {volume} {8}},\ \bibinfo {pages} {565} (\bibinfo {year}
  {2008})}\BibitemShut {NoStop}%
\bibitem [{\citenamefont {Pletikosi\'{c}}\ \emph {et~al.}(2009)\citenamefont
  {Pletikosi\'{c}}, \citenamefont {Kralj}, \citenamefont {Pervan},
  \citenamefont {Brako}, \citenamefont {Coraux}, \citenamefont {N'Diaye} \emph
  {et~al.}}]{Pletikosic2009}%
  \BibitemOpen
  \bibfield  {author} {\bibinfo {author} {\bibfnamefont {I.}~\bibnamefont
  {Pletikosi\'{c}}}, \bibinfo {author} {\bibfnamefont {M.}~\bibnamefont
  {Kralj}}, \bibinfo {author} {\bibfnamefont {P.}~\bibnamefont {Pervan}},
  \bibinfo {author} {\bibfnamefont {R.}~\bibnamefont {Brako}}, \bibinfo
  {author} {\bibfnamefont {J.}~\bibnamefont {Coraux}}, \bibinfo {author}
  {\bibfnamefont {A.~T.}\ \bibnamefont {N'Diaye}},  \emph {et~al.},\ }\href
  {\doibase 10.1103/PhysRevLett.102.056808} {\bibfield  {journal} {\bibinfo
  {journal} {Phys. Rev. Lett.}\ }\textbf {\bibinfo {volume} {102}},\ \bibinfo
  {pages} {056808} (\bibinfo {year} {2009})}\BibitemShut {NoStop}%
\bibitem [{\citenamefont {Kralj}\ \emph {et~al.}(2011)\citenamefont {Kralj},
  \citenamefont {Pletikosi\'{c}}, \citenamefont {Petrovi\'{c}}, \citenamefont
  {Pervan}, \citenamefont {Milun}, \citenamefont {N'Diaye} \emph
  {et~al.}}]{Kralj2011}%
  \BibitemOpen
  \bibfield  {author} {\bibinfo {author} {\bibfnamefont {M.}~\bibnamefont
  {Kralj}}, \bibinfo {author} {\bibfnamefont {I.}~\bibnamefont
  {Pletikosi\'{c}}}, \bibinfo {author} {\bibfnamefont {M.}~\bibnamefont
  {Petrovi\'{c}}}, \bibinfo {author} {\bibfnamefont {P.}~\bibnamefont
  {Pervan}}, \bibinfo {author} {\bibfnamefont {M.}~\bibnamefont {Milun}},
  \bibinfo {author} {\bibfnamefont {A.~T.}\ \bibnamefont {N'Diaye}},  \emph
  {et~al.},\ }\href {\doibase 10.1103/PhysRevB.84.075427} {\bibfield  {journal}
  {\bibinfo  {journal} {Phys. Rev. B}\ }\textbf {\bibinfo {volume} {84}},\
  \bibinfo {pages} {075427} (\bibinfo {year} {2011})}\BibitemShut {NoStop}%
\bibitem [{\citenamefont {Busse}\ \emph {et~al.}(2011)\citenamefont {Busse},
  \citenamefont {Lazi\'{c}}, \citenamefont {Djemour}, \citenamefont {Coraux},
  \citenamefont {Gerber}, \citenamefont {Atodiresei} \emph
  {et~al.}}]{Busse2011a}%
  \BibitemOpen
  \bibfield  {author} {\bibinfo {author} {\bibfnamefont {C.}~\bibnamefont
  {Busse}}, \bibinfo {author} {\bibfnamefont {P.}~\bibnamefont {Lazi\'{c}}},
  \bibinfo {author} {\bibfnamefont {R.}~\bibnamefont {Djemour}}, \bibinfo
  {author} {\bibfnamefont {J.}~\bibnamefont {Coraux}}, \bibinfo {author}
  {\bibfnamefont {T.}~\bibnamefont {Gerber}}, \bibinfo {author} {\bibfnamefont
  {N.}~\bibnamefont {Atodiresei}},  \emph {et~al.},\ }\href {\doibase
  10.1103/PhysRevLett.107.036101} {\bibfield  {journal} {\bibinfo  {journal}
  {Phys. Rev. Lett.}\ }\textbf {\bibinfo {volume} {107}},\ \bibinfo {pages}
  {036101} (\bibinfo {year} {2011})}\BibitemShut {NoStop}%
\bibitem [{\citenamefont {Runte}\ \emph {et~al.}(2014)\citenamefont {Runte},
  \citenamefont {Lazi\'{c}}, \citenamefont {Vo-Van}, \citenamefont {Coraux},
  \citenamefont {Zegenhagen},\ and\ \citenamefont {Busse}}]{Runte2014}%
  \BibitemOpen
  \bibfield  {author} {\bibinfo {author} {\bibfnamefont {S.}~\bibnamefont
  {Runte}}, \bibinfo {author} {\bibfnamefont {P.}~\bibnamefont {Lazi\'{c}}},
  \bibinfo {author} {\bibfnamefont {C.}~\bibnamefont {Vo-Van}}, \bibinfo
  {author} {\bibfnamefont {J.}~\bibnamefont {Coraux}}, \bibinfo {author}
  {\bibfnamefont {J.}~\bibnamefont {Zegenhagen}}, \ and\ \bibinfo {author}
  {\bibfnamefont {C.}~\bibnamefont {Busse}},\ }\href {\doibase
  10.1103/PhysRevB.89.155427} {\bibfield  {journal} {\bibinfo  {journal} {Phys.
  Rev. B}\ }\textbf {\bibinfo {volume} {89}},\ \bibinfo {pages} {155427}
  (\bibinfo {year} {2014})}\BibitemShut {NoStop}%
\bibitem [{\citenamefont {Hattab}\ \emph {et~al.}(2012)\citenamefont {Hattab},
  \citenamefont {N'Diaye}, \citenamefont {Wall}, \citenamefont {Klein},
  \citenamefont {Jnawali}, \citenamefont {Coraux} \emph {et~al.}}]{Hattab2012}%
  \BibitemOpen
  \bibfield  {author} {\bibinfo {author} {\bibfnamefont {H.}~\bibnamefont
  {Hattab}}, \bibinfo {author} {\bibfnamefont {A.~T.}\ \bibnamefont {N'Diaye}},
  \bibinfo {author} {\bibfnamefont {D.}~\bibnamefont {Wall}}, \bibinfo {author}
  {\bibfnamefont {C.}~\bibnamefont {Klein}}, \bibinfo {author} {\bibfnamefont
  {G.}~\bibnamefont {Jnawali}}, \bibinfo {author} {\bibfnamefont
  {J.}~\bibnamefont {Coraux}},  \emph {et~al.},\ }\href {\doibase
  10.1021/nl203530t} {\bibfield  {journal} {\bibinfo  {journal} {Nano Lett.}\
  }\textbf {\bibinfo {volume} {12}},\ \bibinfo {pages} {678} (\bibinfo {year}
  {2012})}\BibitemShut {NoStop}%
\bibitem [{\citenamefont {Coraux}\ \emph {et~al.}(2009)\citenamefont {Coraux},
  \citenamefont {N'Diaye}, \citenamefont {Engler}, \citenamefont {Busse},
  \citenamefont {Wall}, \citenamefont {Buckanie}, \citenamefont {{Meyer Zu
  Heringdorf}} \emph {et~al.}}]{Coraux2009a}%
  \BibitemOpen
  \bibfield  {author} {\bibinfo {author} {\bibfnamefont {J.}~\bibnamefont
  {Coraux}}, \bibinfo {author} {\bibfnamefont {A.~T.}\ \bibnamefont {N'Diaye}},
  \bibinfo {author} {\bibfnamefont {M.}~\bibnamefont {Engler}}, \bibinfo
  {author} {\bibfnamefont {C.}~\bibnamefont {Busse}}, \bibinfo {author}
  {\bibfnamefont {D.}~\bibnamefont {Wall}}, \bibinfo {author} {\bibfnamefont
  {N.}~\bibnamefont {Buckanie}}, \bibinfo {author} {\bibfnamefont {F.~J.}\
  \bibnamefont {{Meyer Zu Heringdorf}}},  \emph {et~al.},\ }\href {\doibase
  10.1088/1367-2630/11/2/023006} {\bibfield  {journal} {\bibinfo  {journal}
  {New J. Phys.}\ }\textbf {\bibinfo {volume} {11}},\ \bibinfo {pages} {023006}
  (\bibinfo {year} {2009})}\BibitemShut {NoStop}%
\bibitem [{\citenamefont {Horcas}\ \emph {et~al.}(2007)\citenamefont {Horcas},
  \citenamefont {Fern\'{a}ndez}, \citenamefont {G\'{o}mez-Rodr\'{\i}guez},
  \citenamefont {Colchero}, \citenamefont {G\'{o}mez-Herrero},\ and\
  \citenamefont {Baro}}]{Horcas2007}%
  \BibitemOpen
  \bibfield  {author} {\bibinfo {author} {\bibfnamefont {I.}~\bibnamefont
  {Horcas}}, \bibinfo {author} {\bibfnamefont {R.}~\bibnamefont
  {Fern\'{a}ndez}}, \bibinfo {author} {\bibfnamefont {J.~M.}\ \bibnamefont
  {G\'{o}mez-Rodr\'{\i}guez}}, \bibinfo {author} {\bibfnamefont
  {J.}~\bibnamefont {Colchero}}, \bibinfo {author} {\bibfnamefont
  {J.}~\bibnamefont {G\'{o}mez-Herrero}}, \ and\ \bibinfo {author}
  {\bibfnamefont {A.~M.}\ \bibnamefont {Baro}},\ }\href {\doibase
  10.1063/1.2432410} {\bibfield  {journal} {\bibinfo  {journal} {Rev. Sci.
  Instrum.}\ }\textbf {\bibinfo {volume} {78}},\ \bibinfo {pages} {013705}
  (\bibinfo {year} {2007})}\BibitemShut {NoStop}%
\bibitem [{\citenamefont {N'Diaye}\ \emph {et~al.}(2008)\citenamefont
  {N'Diaye}, \citenamefont {Coraux}, \citenamefont {Plasa}, \citenamefont
  {Busse},\ and\ \citenamefont {Michely}}]{N'Diaye2008}%
  \BibitemOpen
  \bibfield  {author} {\bibinfo {author} {\bibfnamefont {A.~T.}\ \bibnamefont
  {N'Diaye}}, \bibinfo {author} {\bibfnamefont {J.}~\bibnamefont {Coraux}},
  \bibinfo {author} {\bibfnamefont {T.~N.}\ \bibnamefont {Plasa}}, \bibinfo
  {author} {\bibfnamefont {C.}~\bibnamefont {Busse}}, \ and\ \bibinfo {author}
  {\bibfnamefont {T.}~\bibnamefont {Michely}},\ }\href {\doibase
  10.1088/1367-2630/10/4/043033} {\bibfield  {journal} {\bibinfo  {journal}
  {New J. Phys.}\ }\textbf {\bibinfo {volume} {10}},\ \bibinfo {pages} {043033}
  (\bibinfo {year} {2008})}\BibitemShut {NoStop}%
\bibitem [{\citenamefont {Lu}\ \emph {et~al.}(2009)\citenamefont {Lu},
  \citenamefont {Arroyo},\ and\ \citenamefont {Huang}}]{Lu2009}%
  \BibitemOpen
  \bibfield  {author} {\bibinfo {author} {\bibfnamefont {Q.}~\bibnamefont
  {Lu}}, \bibinfo {author} {\bibfnamefont {M.}~\bibnamefont {Arroyo}}, \ and\
  \bibinfo {author} {\bibfnamefont {R.}~\bibnamefont {Huang}},\ }\href
  {\doibase 10.1088/0022-3727/42/10/102002} {\bibfield  {journal} {\bibinfo
  {journal} {J. Phys. D: Appl. Phys.}\ }\textbf {\bibinfo {volume} {42}},\
  \bibinfo {pages} {102002} (\bibinfo {year} {2009})}\BibitemShut {NoStop}%
\bibitem [{\citenamefont {Starodub}\ \emph {et~al.}(2011)\citenamefont
  {Starodub}, \citenamefont {Bostwick}, \citenamefont {Moreschini},
  \citenamefont {Nie}, \citenamefont {Gabaly}, \citenamefont {McCarty} \emph
  {et~al.}}]{Starodub2011}%
  \BibitemOpen
  \bibfield  {author} {\bibinfo {author} {\bibfnamefont {E.}~\bibnamefont
  {Starodub}}, \bibinfo {author} {\bibfnamefont {A.}~\bibnamefont {Bostwick}},
  \bibinfo {author} {\bibfnamefont {L.}~\bibnamefont {Moreschini}}, \bibinfo
  {author} {\bibfnamefont {S.}~\bibnamefont {Nie}}, \bibinfo {author}
  {\bibfnamefont {F.}~\bibnamefont {Gabaly}}, \bibinfo {author} {\bibfnamefont
  {K.}~\bibnamefont {McCarty}},  \emph {et~al.},\ }\href {\doibase
  10.1103/PhysRevB.83.125428} {\bibfield  {journal} {\bibinfo  {journal} {Phys.
  Rev. B}\ }\textbf {\bibinfo {volume} {83}},\ \bibinfo {pages} {125428}
  (\bibinfo {year} {2011})}\BibitemShut {NoStop}%
\bibitem [{\citenamefont {Okabe}\ \emph {et~al.}(1992)\citenamefont {Okabe},
  \citenamefont {Boots}, \citenamefont {Suhihara},\ and\ \citenamefont
  {Chiu}}]{Okabe-book}%
  \BibitemOpen
  \bibfield  {author} {\bibinfo {author} {\bibfnamefont {A.}~\bibnamefont
  {Okabe}}, \bibinfo {author} {\bibfnamefont {B.}~\bibnamefont {Boots}},
  \bibinfo {author} {\bibfnamefont {K.}~\bibnamefont {Suhihara}}, \ and\
  \bibinfo {author} {\bibfnamefont {S.~N.}\ \bibnamefont {Chiu}},\ }\href@noop
  {} {\emph {\bibinfo {title} {Spatial Tessellations: Concepts and Applications
  of Voronoi Diagrams}}},\ \bibinfo {edition} {2nd}\ ed.\ (\bibinfo
  {publisher} {Wiley},\ \bibinfo {address} {Chichester},\ \bibinfo {year}
  {1992})\BibitemShut {NoStop}%
\bibitem [{\citenamefont {Honda}(1983)}]{Honda}%
  \BibitemOpen
  \bibfield  {author} {\bibinfo {author} {\bibfnamefont {H.}~\bibnamefont
  {Honda}},\ }\href {\doibase 10.1016/S0074-7696(08)62339-6} {\bibfield
  {journal} {\bibinfo  {journal} {Int. Rev. Cytol.}\ }\textbf {\bibinfo
  {volume} {81}},\ \bibinfo {pages} {191} (\bibinfo {year} {1983})}\BibitemShut
  {NoStop}%
\bibitem [{\citenamefont {Hilhorsta}(2008)}]{Hilhorsta2008}%
  \BibitemOpen
  \bibfield  {author} {\bibinfo {author} {\bibfnamefont {H.~J.}\ \bibnamefont
  {Hilhorsta}},\ }\href {\doibase 10.1140/epjb/e2008-00003-7} {\bibfield
  {journal} {\bibinfo  {journal} {Eur. Phys. J. B}\ }\textbf {\bibinfo {volume}
  {64}},\ \bibinfo {pages} {437} (\bibinfo {year} {2008})}\BibitemShut
  {NoStop}%
\bibitem [{\citenamefont {Zhang}\ and\ \citenamefont
  {Arroyo}(2014)}]{Zhang2014}%
  \BibitemOpen
  \bibfield  {author} {\bibinfo {author} {\bibfnamefont {K.}~\bibnamefont
  {Zhang}}\ and\ \bibinfo {author} {\bibfnamefont {M.}~\bibnamefont {Arroyo}},\
  }\href {\doibase 10.1016/j.jmps.2014.07.012} {\bibfield  {journal} {\bibinfo
  {journal} {J. Mech. Phys. Solids}\ }\textbf {\bibinfo {volume} {72}},\
  \bibinfo {pages} {61} (\bibinfo {year} {2014})}\BibitemShut {NoStop}%
\bibitem [{\citenamefont {Robinson}\ \emph {et~al.}(2009)\citenamefont
  {Robinson}, \citenamefont {Puls}, \citenamefont {Staley}, \citenamefont
  {Stitt}, \citenamefont {Fanton}, \citenamefont {Emtsev} \emph
  {et~al.}}]{Robinson2009}%
  \BibitemOpen
  \bibfield  {author} {\bibinfo {author} {\bibfnamefont {J.~A.}\ \bibnamefont
  {Robinson}}, \bibinfo {author} {\bibfnamefont {C.~P.}\ \bibnamefont {Puls}},
  \bibinfo {author} {\bibfnamefont {N.~E.}\ \bibnamefont {Staley}}, \bibinfo
  {author} {\bibfnamefont {J.~P.}\ \bibnamefont {Stitt}}, \bibinfo {author}
  {\bibfnamefont {M.~A.}\ \bibnamefont {Fanton}}, \bibinfo {author}
  {\bibfnamefont {K.~V.}\ \bibnamefont {Emtsev}},  \emph {et~al.},\ }\href
  {\doibase 10.1021/nl802852p} {\bibfield  {journal} {\bibinfo  {journal} {Nano
  Lett.}\ }\textbf {\bibinfo {volume} {9}},\ \bibinfo {pages} {964} (\bibinfo
  {year} {2009})}\BibitemShut {NoStop}%
\bibitem [{\citenamefont {Zhang}\ and\ \citenamefont
  {Arroyo}(2013)}]{Zhang2013}%
  \BibitemOpen
  \bibfield  {author} {\bibinfo {author} {\bibfnamefont {K.}~\bibnamefont
  {Zhang}}\ and\ \bibinfo {author} {\bibfnamefont {M.}~\bibnamefont {Arroyo}},\
  }\href {\doibase 10.1063/1.4804265} {\bibfield  {journal} {\bibinfo
  {journal} {J. Appl. Phys.}\ }\textbf {\bibinfo {volume} {113}},\ \bibinfo
  {pages} {193501} (\bibinfo {year} {2013})}\BibitemShut {NoStop}%
\bibitem [{\citenamefont {Cleymand}\ \emph {et~al.}(2001)\citenamefont
  {Cleymand}, \citenamefont {Coupeau},\ and\ \citenamefont
  {Grilh\'{e}}}]{Cleymand2001}%
  \BibitemOpen
  \bibfield  {author} {\bibinfo {author} {\bibfnamefont {F.}~\bibnamefont
  {Cleymand}}, \bibinfo {author} {\bibfnamefont {C.}~\bibnamefont {Coupeau}}, \
  and\ \bibinfo {author} {\bibfnamefont {J.}~\bibnamefont {Grilh\'{e}}},\
  }\href {\doibase 10.1016/S1359-6462(01)00969-1} {\bibfield  {journal}
  {\bibinfo  {journal} {Scr. Mater.}\ }\textbf {\bibinfo {volume} {44}},\
  \bibinfo {pages} {2623} (\bibinfo {year} {2001})}\BibitemShut {NoStop}%
\bibitem [{\citenamefont {Vella}\ \emph {et~al.}(2009)\citenamefont {Vella},
  \citenamefont {Bico}, \citenamefont {Boudaoud}, \citenamefont {Roman},\ and\
  \citenamefont {Reis}}]{Vella2009}%
  \BibitemOpen
  \bibfield  {author} {\bibinfo {author} {\bibfnamefont {D.}~\bibnamefont
  {Vella}}, \bibinfo {author} {\bibfnamefont {J.}~\bibnamefont {Bico}},
  \bibinfo {author} {\bibfnamefont {A.}~\bibnamefont {Boudaoud}}, \bibinfo
  {author} {\bibfnamefont {B.}~\bibnamefont {Roman}}, \ and\ \bibinfo {author}
  {\bibfnamefont {P.~M.}\ \bibnamefont {Reis}},\ }\href {\doibase
  10.1073/pnas.0902160106} {\bibfield  {journal} {\bibinfo  {journal} {Proc.
  Natl. Acad. Sci. U. S. A.}\ }\textbf {\bibinfo {volume} {106}},\ \bibinfo
  {pages} {10901} (\bibinfo {year} {2009})}\BibitemShut {NoStop}%
\bibitem [{\citenamefont {Aitken}\ and\ \citenamefont
  {Huang}(2010)}]{Aitken2010}%
  \BibitemOpen
  \bibfield  {author} {\bibinfo {author} {\bibfnamefont {Z.~H.}\ \bibnamefont
  {Aitken}}\ and\ \bibinfo {author} {\bibfnamefont {R.}~\bibnamefont {Huang}},\
  }\href {\doibase 10.1063/1.3437642} {\bibfield  {journal} {\bibinfo
  {journal} {J. Appl. Phys.}\ }\textbf {\bibinfo {volume} {107}},\ \bibinfo
  {pages} {123531} (\bibinfo {year} {2010})}\BibitemShut {NoStop}%
\bibitem [{\citenamefont {Timoshenko}\ and\ \citenamefont
  {Woinowsky-Krieger}(1959)}]{Timoshenko-book}%
  \BibitemOpen
  \bibfield  {author} {\bibinfo {author} {\bibfnamefont {S.}~\bibnamefont
  {Timoshenko}}\ and\ \bibinfo {author} {\bibfnamefont {S.}~\bibnamefont
  {Woinowsky-Krieger}},\ }\href@noop {} {\emph {\bibinfo {title} {Theory of
  Plates and Shells}}},\ \bibinfo {edition} {2nd}\ ed.\ (\bibinfo  {publisher}
  {McGraw-Hill},\ \bibinfo {address} {New York},\ \bibinfo {year}
  {1959})\BibitemShut {NoStop}%
\end{thebibliography}%

\end{document}